\documentclass[twocolumn]{aastex6}

\usepackage{siriodef}
\usepackage{capt-of}
\usepackage{graphicx}
\usepackage{subfigure}
\usepackage{amsmath}

\newcommand{\Sq}{\ensuremath{S_\mathrm{Q}}}
\newcommand{\Cq}{\ensuremath{C_\mathrm{Q}}}

\defcitealias{belli17mosfire}{Paper I}

\shorttitle{Spectroscopy of Quiescent Galaxies at $1.5 < z < 2.5$ - II. Stellar Ages}
\shortauthors{Belli, Newman and Ellis}
\submitted{Accepted for publication in ApJ}

\begin{document}

\title{MOSFIRE Spectroscopy of Quiescent Galaxies at $1.5 < \lowercase{z} < 2.5$. \\
II. Star Formation Histories and Galaxy Quenching}

\author{Sirio Belli\altaffilmark{1}, Andrew B. Newman\altaffilmark{2}, Richard S. Ellis\altaffilmark{3}}
\altaffiltext{1}{Max-Planck-Institut f\"ur Extraterrestrische Physik (MPE), Giessenbachstr. 1, D-85748 Garching, Germany}
\altaffiltext{2}{The Observatories of the Carnegie Institution for Science, 813 Santa Barbara St., Pasadena, CA 91101, USA}
\altaffiltext{3}{Department of Physics and Astronomy, University College London, Gower Street, London WC1E 6BT, UK}


\begin{abstract}

We investigate the stellar populations for a sample of 24 quiescent galaxies at $1.5 < z < 2.5$ using deep rest-frame optical spectra obtained with Keck MOSFIRE. By fitting templates simultaneously to the spectroscopic and photometric data, and exploring a variety of star formation histories, we obtain robust measurements of median stellar ages and residual levels of star formation. After subtracting the stellar templates, the stacked spectrum reveals the \Halpha\ and \NII\ emission lines, providing an upper limit on the ongoing star formation rate of $0.9 \pm 0.1$ \Msun/yr. By combining the MOSFIRE data to our sample of Keck LRIS spectra at lower redshift, we analyze in a consistent manner the quiescent population at $1 < z < 2.5$. We find a tight relation (with a scatter of 0.13 dex) between the stellar age and the rest-frame $U-V$ and $V-J$ colors, which can be used to estimate the age of quiescent galaxies given their colors. Applying this age--color relation to large, photometric samples, we are able to model the number density evolution for quiescent galaxies of various ages. We find evidence for two distinct quenching paths: a fast quenching that produces compact post-starburst systems, and a slow quenching of larger galaxies. Fast quenching accounts for about a fifth of the growth of the red sequence at $z\sim1.4$, and half at $z\sim2.2$. We conclude that fast quenching is triggered by dramatic events such as gas-rich mergers, while slow quenching is likely caused by a different physical mechanism.

\end{abstract}



\section{Introduction}
\label{sec:introduction}

The physical mechanism responsible for quenching the star formation activity in massive galaxies remains one of the most important missing pieces in the puzzle of galaxy formation. The existence of a population of massive quiescent galaxies at $z\sim2$ \citep{franx03, cimatti04}, when the universe was only a few Gyr old, requires a remarkably rapid and efficient quenching. These galaxies are also found to be significantly more compact than their local counterparts \citep[e.g.,][]{vandokkum08, newman12}, thus raising additional questions about the process that led to their formation.

Moreover, it is not clear what these systems looked like \emph{before} being quenched, and constraining the properties of their star-forming progenitors is currently a major observational goal. The physical mechanism responsible for quenching and the properties of the progenitors are tightly connected, and represent two aspects of the same open question. A number of possibilities have been put forward, including evolutionary links with submillimeter galaxies \citep{tacconi08, toft14} and compact star-forming galaxies \citep{barro13, vandokkum15}; however, observations are currently unable to discriminate among different scenarios.

A powerful method to study the past evolution of massive quiescent galaxies is the reconstruction of their star formation history (SFH), which can be done by fitting models to the observed spectral energy distribution (SED) or, when available, spectroscopy \citep[see, e.g.,][]{renzini06,conroy13review}. This \emph{archaeological} approach has allowed detailed studies of the local population, revealing an anti-hierarchical assembly in which the most massive objects formed earlier and on shorter timescales \citep[e.g.,][]{thomas05}. Additionally, the stellar metallicity can be measured from high-quality spectra: the state-of-the-art analysis of local quiescent galaxies includes measured abundances for 16 different chemical elements \citep{conroy14}.

The archaeological approach, however, presents significant limitations when used to understand the early evolution of quiescent galaxies. First, the spectra of old stellar populations present very little evolution, a fact that produces the remarkable homogeneity of the local red sequence \citep{bower92}. As a result, ages become increasingly uncertain for older systems. Secondly, there is a degeneracy between the mass \emph{formation} history and the mass \emph{assembly} history. Even when a reliable SFH is found, it is not possible to know whether the various episodes were due to \emph{in-situ} star formation or took place in separate systems that eventually merged together. Being able to directly probe these different scenarios is fundamental for our understanding of galaxy formation.

The limitations of galaxy archaeology can be overcome by observing galaxies at high redshift, where stellar ages are bounded by the age of the universe at the epoch of observation. Rest-frame optical spectra are particular powerful when probing stellar populations younger than about 3 Gyr, due to the presence of Balmer absorption lines that are strong and evolve rapidly with age. In some sense, measuring ages and SFHs is increasingly easier at higher redshift; the difficulty lies in obtaining deep rest-frame optical spectra for faint, distant targets.

Recent studies of large, high-quality spectroscopic samples at intermediate redshift ($z<1$) have shown that the metallicities and stellar ages of quiescent galaxies are consistent with passive evolution over the last 7 Gyr of cosmic history, and that more massive galaxies tend to have older stellar populations \citep{choi14, gallazzi14, wu18, chauke18}, in agreement with the results of kinematic studies \citep{treu05, vanderwel05}. Extending this type of analysis to earlier epochs, when the majority of massive galaxies formed, is observationally challenging. At $1 < z < 1.5$ the use of red-sensitive CCDs enables the collection of relatively large samples \citep{bezanson13,belli15}, but at $z>1.5$ near-infrared observations are needed, and most studies are based on stacked spectra \citep{whitaker13, newman14, onodera15, mendel15, fumagalli16} or on sample sizes of one to a few absorption-line spectra \citep{vandesande13, kriek16, glazebrook17, schreiber18}.
These observations revealed that the typical ages of quiescent galaxies at $z\sim2$ are 1-2 Gyr, with a relatively large age spread at a given redshift, confirming that this is the epoch when the red sequence first formed.

In this work we analyze the SFHs for a sample of 24 quiescent galaxies at $1.5 < z < 2.5$ for which we obtained deep near-infrared observations with the MOSFIRE instrument at Keck. We have already presented this sample in \citet{belli17mosfire}, hereafter \citetalias{belli17mosfire}, where we used the absorption-line spectra to derive the stellar kinematics. In the present work we constrain the SFHs of these galaxies by fitting templates simultaneously to the spectroscopic and the photometric data. This technique enables us to take advantage of both the wide wavelength range of photometric data and the age-sensitive features in the rest-frame optical spectra, and was already successfully applied to our Keck/LRIS sample at $1 < z < 1.5$ in \citet{belli15}.

In addition to the limited availability of deep rest-frame optical spectra, the measurement of stellar ages at high redshift suffers from systematics due to the assumptions involved in the spectral fitting. A functional form for the SFH is often assumed, but the impact of this procedure on the age measurement has not been explored in detail. In this study we attempt a comparison of different SFH models, with the goal of understanding which physical properties are robust against different SFH assumptions.

We briefly present the spectroscopic data in Section~\ref{sec:data}. In Section~\ref{sec:fit} we discuss the spectral fits, perform a detailed comparison between different SFH models, and measure the amount of recent star formation using the \Halpha\ emission line. We then use the results of the spectral fits to calibrate the relation between age and rest-frame color in Section~\ref{sec:age-color}. In Section~\ref{sec:numdens} we apply this relation to the large photometric catalog from the UltraVISTA survey, and we model in detail the number density evolution of the quiescent population. The main result is the identification of two distinct quenching channels, which we discuss in detail in Section~\ref{sec:quenching}. Finally, we summarize our results in Section~\ref{sec:summary}.

Throughout the paper we use AB magnitudes and assume a $\Lambda$CDM cosmology with $\Omega_M$=0.3, $\Omega_{\Lambda}$=0.7 and $H_0$= 70 km s$^{-1}$ Mpc$^{-1}$.


\section{Data}
\label{sec:data}

The present work is based on a sample of 24 deep spectra obtained at Keck with the near-infrared spectrograph MOSFIRE. In this section we briefly present the sample; for a detailed description of the target selection, observations, and data reduction we refer to \citetalias{belli17mosfire}.

We selected massive galaxies with photometric redshift in the range $1.5 < z < 2.5$ from the 3D-HST catalog \citep{momcheva16}, prioritizing systems that are quiescent according to their rest-frame $UVJ$ colors. Since all the targets are in the CANDELS fields \citep{grogin11, koekemoer11}, a wealth of ancillary photometric and imaging data are publicly available. The MOSFIRE observations were carried out in the $Y$ and $J$ bands with exposure times between 4 and 9 hours per pointing. Additional, shorter observations (1 to 2.5 hours) were obtained in the $H$ and $K$ band in order to target the \Halpha\ emission line. A standard data reduction, which includes telluric correction and absolute flux calibration, was performed.

To explore the evolution of quiescent galaxies over a wide redshift range, in Section~\ref{sec:age-color} we will augment the sample with the galaxies at $1 < z < 1.5$ observed with Keck/LRIS and presented in \citet{newman10} and \citet{belli14lris}. This sample is well matched to the MOSFIRE one in terms of target selection and rest-frame wavelength range. As we have shown in our previous studies, our two Keck samples probe the population of massive quiescent galaxies at high redshift in an unbiased way, except for a slight lack of old and dusty objects at $z>1.5$ \citepalias{belli17mosfire}.

The combined Keck sample represents the largest homogeneous set of deep absorption-line spectroscopy of quiescent galaxies at $z>1$. We publicly release the reduced spectra for the entire sample\footnote{The spectra are available as a Contributed Dataset in the Keck Observatory Archive (\url{https://koa.ipac.caltech.edu/Datasets/})}.


\section{Stellar Populations}
\label{sec:fit}

\subsection{Spectral Fitting}
\label{sec:pyspecfit}

We analyze the data using \texttt{pyspecfit} \citep{newman14}, a Bayesian code that fits stellar population templates to the observed spectroscopy and photometry of a galaxy. The code employs the nested sampling algorithm \citep{skilling04} to explore the multidimensional parameter space in an efficient way.

Our spectroscopic data typically cover the rest-frame wavelength range 3700--4200\AA, but the exact interval depends on redshift and target position on the MOSFIRE slitmask. Before the fit, in each spectrum we mask out the pixels at the expected position of the \OII\ emission line, since our templates are purely stellar and do not include the contribution of ionized gas. To reduce the impact of sky lines on the spectrum, we also exclude from the fit the spectral pixels with the strongest sky emission, as measured by the observed sky spectrum.

We take the photometric data from the 3D-HST catalog, which contains publicly available measurements from the UV to the near-infrared. To avoid potential contamination from dust emission, we mask out the IRAC 8\micron\ channel. To account for systematic effects, we add in quadrature a 5\%\ contribution to the uncertainty of each photometric point. Data points corresponding to non-detections are included in the fit with their formal uncertainty.

One limitation of optical and near-infrared spectra is the large uncertainty on the absolute flux calibration, which can be affected by slit loss and variable weather conditions. Instrumental effects can also change the overall shape of the spectral response, particularly near the limits of the observed wavelength range. Since the absolute calibration of photometric measurements is typically much more reliable, during the fit we change the overall shape of the observed spectrum, using a fourth degree polynomial, so that it agrees with the shape given by the broadband photometry. This procedure does not affect any of the spectral features of interest, which extend over narrow spectral ranges. We also scale the error spectra to ensure that the spectroscopic and the photometric data are given appropriate weighting in the fit. To do this, for each galaxy we run a first fit with the sole purpose of calculating the chi-square, which we then use to rescale the error spectrum. The scaling factors are in the range 1.7--2.3, suggesting that the uncertainty on the spectra that is produced by the data reduction is indeed too small.

We generate the synthetic spectra using the \citet{bruzual03} library, and adopting a \citet{chabrier03} initial mass function (IMF) and the dust extinction law from \citet{calzetti00}. A simple stellar population (SSP) of a given age can be used to generate a realistic spectrum once the following four parameters are specified: redshift, stellar velocity dispersion, dust attenuation $A_V$, and stellar metallicity $Z$. We leave these four parameters free during the fit, setting their priors in the following way. The spectroscopic redshift can be determined by eye with reasonable accuracy, and we use a narrow Gaussian prior centered on the values reported in \citetalias{belli17mosfire}. We adopt a Gaussian prior also for the stellar metallicity, centered on the solar value ($Z=0.02$) but wide enough to encompass all reasonable values ($\sigma_Z = 0.005$). For the other parameters we use uninformative priors in the form of uniform distributions: between 50 and 500 km/s for the velocity dispersion, and between 0 and 4 for $A_V$.

An additional free parameter is the age of the stellar population, which means that a model can be fully specified using five parameters. However, the assumption of SSP is likely too simplistic for real galaxies, and more complex SFHs, with a higher number of free parameters, are often considered. We attempt here a systematic exploration of different SFH models, with a twofold goal: first, we want to derive the stellar population properties in a robust way, ensuring that the results are not affected by the assumed form of the SFH; secondly, we are going to assess whether some of the SFH models are better than others in describing the spectra of real galaxies.

Each SFH model is simply a function of lookback time that is normalized to unity, and depends on one or more parameters. The SFH models used in our analysis, together with the priors for each parameter, are listed in Table~\ref{tab:SFH}. We use the following six SFH models:
\begin{itemize}
\item The SSP represents the basic form of SFH, in which all stars form in an instantaneous burst. The only free parameter is the age of the population, for which we adopt a log-uniform prior that is truncated at the age of the universe at the observed redshift.
\item Next, we consider a top-hat model, i.e. a burst with a finite duration. This model has two free parameters: the age and the duration of the burst.
\item The most popular SFH choice is an exponentially declining function, also called \emph{tau model}. This SFH has two free parameters: the age and the timescale $\tau$, and the star formation rate is proportional to $\exp(-t/\tau)$. In this SFH the current star formation rate is lower than at any time in the past, which is a reasonable assumption for galaxies that are, by selection, quiescent.
\item Another popular choice is the delayed tau model, which has the form $t \exp(-t/\tau)$. This SFH has been proposed in order to overcome the unrealistically sharp peak at early times featured by the tau model, without introducing additional free parameters.
\item In order to allow more flexibility, we introduce a new model which consists of a constant SFH followed by an exponential decay, hereafter the \emph{constant + tau model}. This model has three free parameters: the age, the duration of the flat phase, and the timescale $\tau$ for the exponential phase, and can be considered as a generalization of both the top-hat and the tau model. Its main advantage is that it allows the quenching phase to be completely detached from the initial star-forming phase.
\item Finally, we consider a non-parametric SFH with a large degree of flexibility, in order to explore further possibilities such as secondary bursts or rejuvenation episodes. We adopt a piecewise description of the SFH with seven nodes at fixed, logarithmically spaced lookback times (0, 0.25, 0.45, 0.8, 1.4, 2.5, and 4 Gyr). Since the SFH models must be normalized, this corresponds to six free parameters. We choose to leave the amplitude of the first six nodes free, while the amplitude of the oldest node is arbitrarily fixed to unity. To avoid biasing the results we keep the oldest node fixed at 4 Gyr for all galaxies, although this means that, according to the redshift, stars slightly older than the age of the universe may be allowed. The SFH is then constructed by linearly interpolating between the nodes.  We adopt a log-normal prior for each amplitude, centered on 1 and with a width of 2 dex, thus allowing a large amount of freedom.
\end{itemize}

\begin{figure*}[htbp]
\centering
\includegraphics[width=0.9\textwidth]{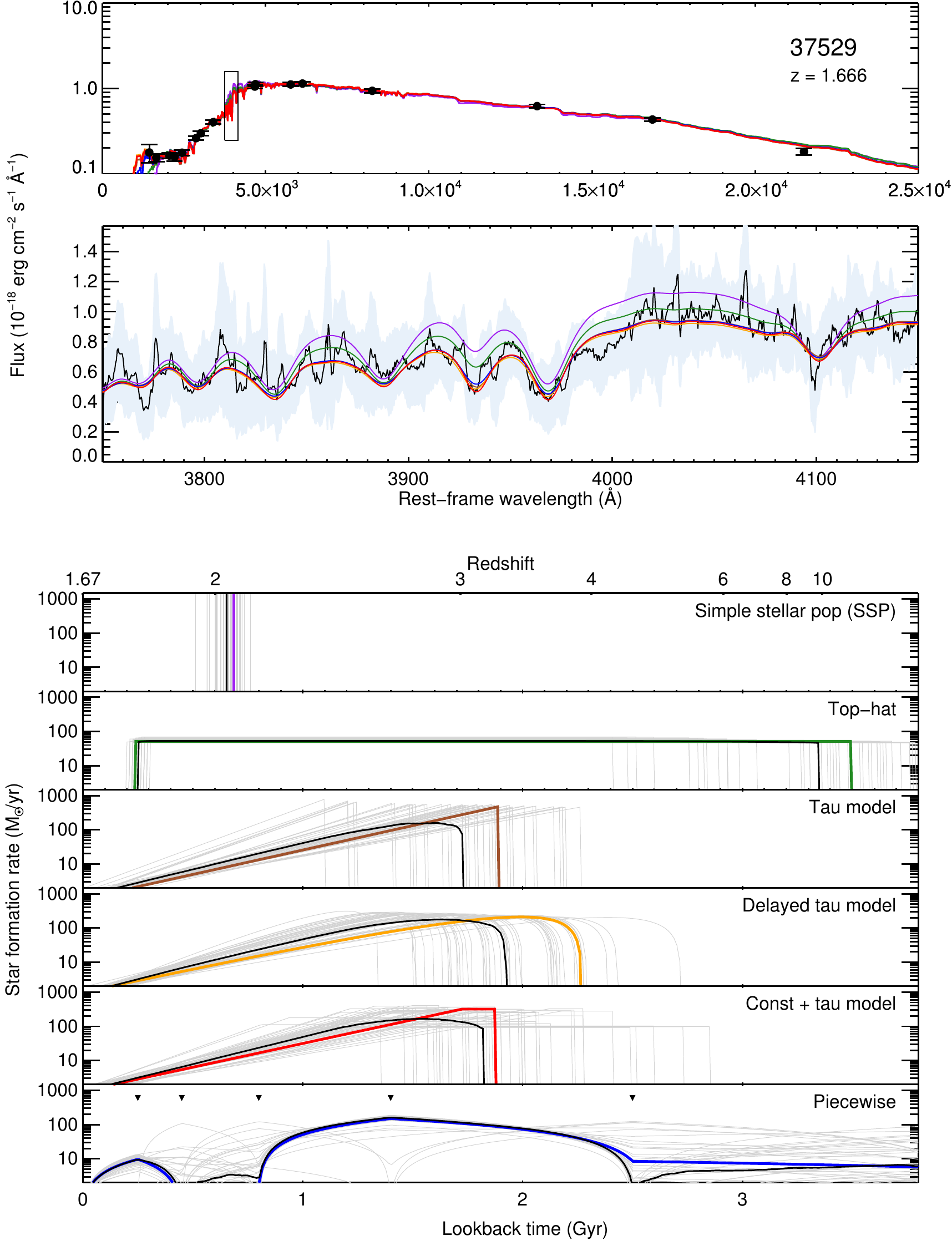}
\caption{Example of spectral fitting for the object 37529, at $z=1.666$. The observed photometry and spectroscopy are fit using six different models for the SFH. Top: observed broadband photometry (black points) and best-fit template for each of the six models (colored lines). The black box indicates the part of the spectrum observed with MOSFIRE. Center: The observed MOSFIRE spectrum (black line) is shown with its uncertainty (shaded region); the same six best-fit templates are also shown (colored lines). Bottom: in each panel, the best-fit SFH is shown as a colored line, where the colors correspond to the best-fit spectra shown in the top two panels; a random subset of SFHs drawn from the posterior distribution is shown in gray; the median SFH is shown in black. The lookback time for this particular galaxy is converted into the corresponding redshift in the top $x$ axis. In the last panel, the nodes of the piecewise SFH are marked by black triangles.}
\label{fig:spectrum_example}
\end{figure*}

\begin{deluxetable}{lll}
\tabletypesize{\footnotesize}
\tablewidth{0pc}
\tablecaption{Priors on the SFH Model Parameters \label{tab:SFH}}
\tablehead{
\colhead{SFH model} & \colhead{Parameters} & \colhead{Prior}
}
\startdata
SSP             & age       & LogUnif(100 Myr, $t_{U}$)    \vspace*{6pt} \\
Top-hat         & age       & LogUnif(100 Myr, $t_{U}$)     \\
                & duration  & LogUnif(100 Myr, $t_{U}$)    \vspace*{6pt} \\
Tau             & age       & LogUnif(100 Myr, $t_{U}$)     \\
                & $\tau$    & LogUnif(10 Myr, 10 Gyr)          \vspace*{6pt} \\
Delayed tau     & age       & LogUnif(100 Myr, $t_{U}$)     \\
                & $\tau$    & LogUnif(10 Myr, 10 Gyr)          \vspace*{6pt} \\
Const + tau  & age          & LogUnif(100 Myr, $t_{U}$)     \\
                & duration  & LogUnif(100 Myr, $t_{U}$)     \\
                & $\tau$    & LogUnif(10 Myr, 10 Gyr)          \vspace*{6pt} \\
Piecewise\tablenotemark{a}       & $A_0$      & LogNormal(0, 2)     \\
                & $A_{0.25}$    & LogNormal(0, 2)     \\
                & $A_{0.45}$    & LogNormal(0, 2)     \\
                & $A_{0.8}$    & LogNormal(0, 2)     \\
                & $A_{1.4}$    & LogNormal(0, 2)     \\
                & $A_{2.5}$    & LogNormal(0, 2)     \\
\enddata
\tablecomments{$t_{U}$ is the age of the universe at the spectroscopic redshift of the galaxy. LogUnif($a, b$) is the log-uniform distribution between $a$ and $b$. LogNormal($\mu, \sigma$) is the log-normal distribution, such that the logarithm of the variable has mean $\mu$ and standard deviation $\sigma$.}
\tablenotetext{a}{$A_X$ is the amplitude of the SFH at a lookback time of $X$ Gyr; $A_4$ is fixed to 1.}
\end{deluxetable}

The SFH represents the \emph{relative} contribution of stars of different ages to a population. The total luminosity is then determined by the total number of stars in the population, i.e., by the total stellar mass. For each template, this can be calculated by multiplying its mass-to-light ratio by the observed luminosity, corrected for cosmological dimming. The stellar mass is therefore an output of the fitting procedure, but does not represent a free parameter of the model. The same is true for the instantaneous star formation rate, which can be easily derived from the SFH after accounting for the mass loss due to winds and supernovae. We make the simplifying assumption of instantaneous recycling, so that a fixed return fraction $R$ of the newly formed stars is immediately lost and does not contribute to the stellar mass. For a Chabrier IMF the return fraction is $R=0.41$ \citep{madau14}. The relation between the SFH and the star formation rate (SFR) at any cosmic time $t$ is therefore:
\begin{equation}
  \label{eq:SFR}
  \mathrm{SFH}(t) \cdot \Mstar = \frac{d\Mstar(t)}{dt} = (1-R) \cdot \mathrm{SFR}(t) \; .
\end{equation}

\subsection{Results of the Spectral Fitting}

\begin{figure*}[htbp]
\centering
\includegraphics[height=0.9\textheight]{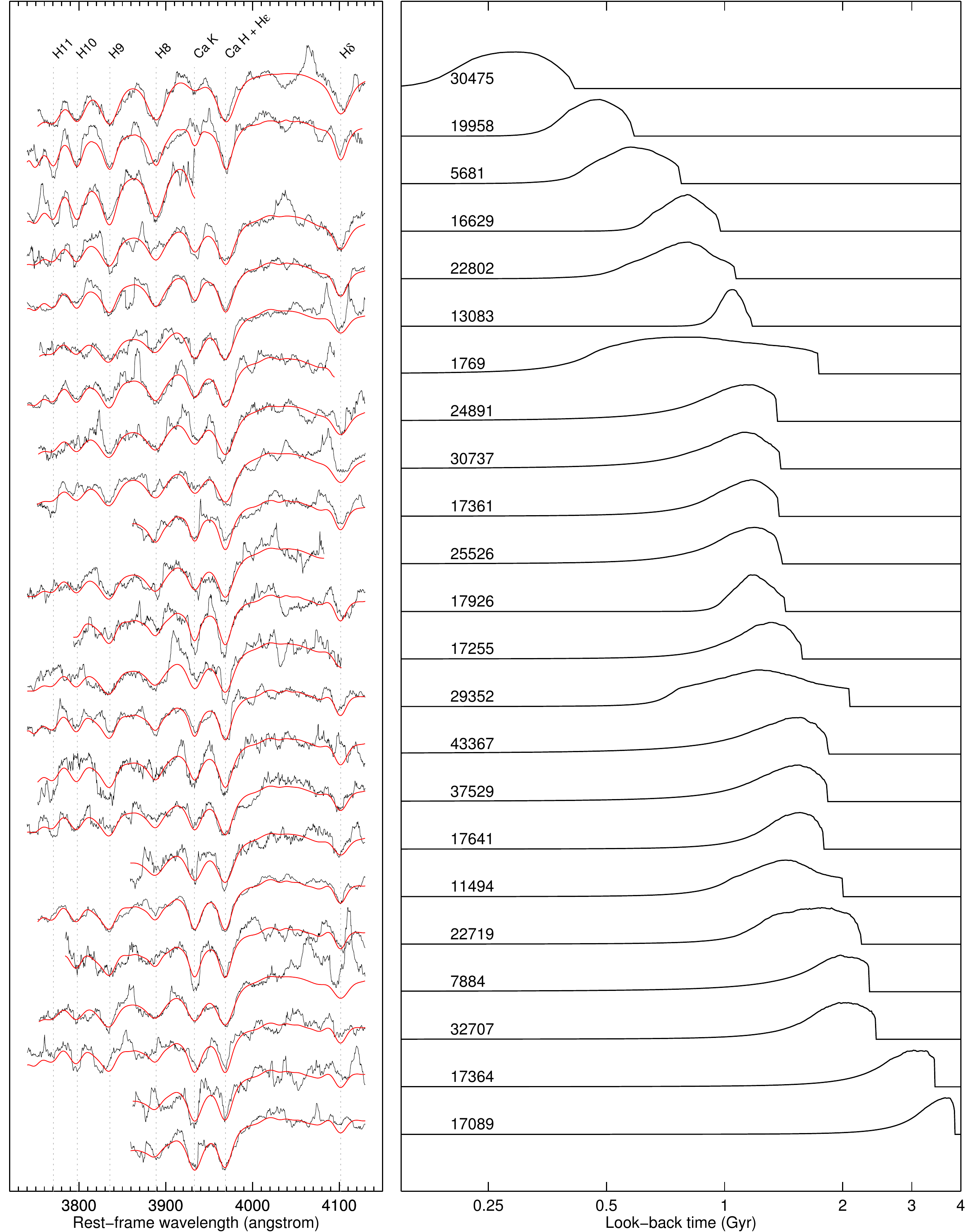}
\caption{Rest-frame optical spectra and derived SFHs for the galaxies in the MOSFIRE sample, sorted by increasing median age. Left: for each galaxy, the MOSFIRE spectrum (black) and the best-fit template (red) are shown; the most important absorption lines are indicated by vertical dashed lines and labeled at the top. Right: the median SFH assuming a constant + tau model is shown for each object.}
\label{fig:sample}
\end{figure*}

Figure \ref{fig:spectrum_example} illustrates the spectral fitting for one representative galaxy drawn from our sample. The top two panels show, in black, the observed photometry and the MOSFIRE spectrum. Six colored lines, corresponding to the best-fit templates for the six types of SFH, are overplotted. For each SFH model the colored line in the top panel is exactly the same as the one in the second panel, because we fit simultaneously the spectroscopic and the photometric data. In the bottom half of the figure, the six types of SFHs are plotted in units of star formation rate, using Equation \ref{eq:SFR} for the conversion. In each panel, the best-fit SFH is shown as a colored line, while thin gray lines illustrate an even sampling of the posterior distribution. We also construct the median SFH by taking, at each lookback time, the median of the gray lines. This is shown as a thick black line and is more representative of the posterior distribution than the single best-fit curve. Throughout this work, when we quote the results of spectral fitting such as ages, star formation rates, etc, we always use the median calculated from an even sampling of the posterior distribution, and not the best-fit value.

\begin{deluxetable*}{llccccccl}
\tabletypesize{\footnotesize}
\tablewidth{0pc}
\tablecaption{Stellar Population Properties of the MOSFIRE Sample \label{tab:sample}}
\tablehead{
\colhead{ID} & \colhead{Field} & \colhead{z} & \colhead{$t_{50}$\tablenotemark{a}} &  \colhead{$A_V$\tablenotemark{a}} &  \colhead{$Z$\tablenotemark{a}} &  \colhead{log \Mstar/\Msun\tablenotemark{a}} &  \colhead{log sSFR\tablenotemark{a,b}} & \colhead{SFR(\Halpha)\tablenotemark{c}}
\\
(3D-HST) & & & $10^9$ yr & mag & & & yr$^{-1}$ & \Msun/yr
}
\startdata
30475\tablenotemark{d} & UDS & 1.63 & 0.30   $ \pm $ 0.06 & 0.85 $ \pm $ 0.10 & 0.013 $ \pm $ 0.004 & 10.71 $ \pm $ 0.02 & -11.0 $ \pm $ 1.0 &  $<$ 8.9 \\
19958 & COSMOS & 1.72 & 0.48$ \pm $ 0.04 & 0.73 $ \pm $ 0.06 & 0.012 $ \pm $ 0.002 & 10.64 $ \pm $ 0.01 & -14.2 $ \pm $ 3.0 &  \nodata  \\
5681 & COSMOS & 2.43 & 0.61 $ \pm $ 0.07 & 0.64 $ \pm $ 0.09 & 0.017 $ \pm $ 0.003 & 10.91 $ \pm $ 0.02 & -12.9 $ \pm $ 3.6 &  $<$ 14.9 \\
16629 & COSMOS & 1.66 & 0.81$ \pm $ 0.07 & 0.44 $ \pm $ 0.05 & 0.017 $ \pm $ 0.003 & 10.59 $ \pm $ 0.01 & -17.0 $ \pm $ 5.7 &  $<$ 7.2 \\
22802 & UDS & 1.67 & 0.83   $ \pm $ 0.1 & 0.62  $ \pm $ 0.09 & 0.015 $ \pm $ 0.004 & 11.01 $ \pm $ 0.02 & -13.2 $ \pm $ 4.9 &  $<$ 6.8 \\
13083 & COSMOS & 2.09 & 1.1 $ \pm $ 0.1 & 0.20  $ \pm $ 0.10 & 0.022 $ \pm $ 0.003 & 10.95 $ \pm $ 0.03 & -20.0 $ \pm $ 9.8 &  $<$ 14.5 \\
1769 & COSMOS & 2.30 & 1.1  $ \pm $ 0.2 & 0.43  $ \pm $ 0.10 & 0.022 $ \pm $ 0.005 & 11.16 $ \pm $ 0.03 & -11.1 $ \pm $ 0.2 &  \nodata  \\
24891 & UDS & 1.60 & 1.1    $ \pm $ 0.1 & 0.55  $ \pm $ 0.10 & 0.015 $ \pm $ 0.004 & 10.87 $ \pm $ 0.02 & -10.4 $ \pm $ 0.09 &  $<$ 6.3 \\
30737\tablenotemark{d} & UDS & 1.62 & 1.1    $ \pm $ 0.1 & 0.51  $ \pm $ 0.09 & 0.019 $ \pm $ 0.004 & 11.23 $ \pm $ 0.02 & -10.6 $ \pm $ 0.08 &  $<$ 15.1 \\
17361 & COSMOS & 1.53 & 1.1 $ \pm $ 0.09 & 0.37 $ \pm $ 0.07 & 0.011 $ \pm $ 0.002 & 10.74 $ \pm $ 0.02 & -10.9 $ \pm $ 0.1 &  $<$ 1.7 \\
25526 & EGS & 1.75 & 1.2    $ \pm $ 0.2 & 0.34  $ \pm $ 0.10 & 0.015 $ \pm $ 0.005 & 10.69 $ \pm $ 0.03 & -10.9 $ \pm $ 1.3 &  \nodata  \\
17926 & EGS & 1.57 & 1.2    $ \pm $ 0.1 & 0.33  $ \pm $ 0.09 & 0.012 $ \pm $ 0.002 & 10.85 $ \pm $ 0.05 & -20.1 $ \pm $ 10.6 &  $<$ 2.0 \\
17255 & COSMOS & 1.74 & 1.3 $ \pm $ 0.2 & 0.31  $ \pm $ 0.06 & 0.017 $ \pm $ 0.004 & 10.81 $ \pm $ 0.03 & -11.3 $ \pm $ 0.4 &  \nodata  \\
29352 & UDS & 1.69 & 1.5    $ \pm $ 0.3 & 0.23  $ \pm $ 0.07 & 0.012 $ \pm $ 0.003 & 10.82 $ \pm $ 0.05 & -12.6 $ \pm $ 5.5 &  \nodata  \\
43367 & UDS & 1.62 & 1.5    $ \pm $ 0.3 & 0.68  $ \pm $ 0.10 & 0.017 $ \pm $ 0.004 & 11.06 $ \pm $ 0.04 & -10.4 $ \pm $ 0.1 &  $<$ 9.2 \\
37529 & UDS & 1.67 & 1.5    $ \pm $ 0.2 & 0.58  $ \pm $ 0.07 & 0.014 $ \pm $ 0.004 & 10.95 $ \pm $ 0.04 & -10.8 $ \pm $ 0.08 &  $<$ 7.9 \\
17641 & COSMOS & 1.53 & 1.5 $ \pm $ 0.2 & 0.28  $ \pm $ 0.05 & 0.013 $ \pm $ 0.003 & 10.61 $ \pm $ 0.04 & -11.5 $ \pm $ 1.8 &  $<$ 3.0 \\
11494 & COSMOS & 2.09 & 1.6 $ \pm $ 0.2 & 0.26  $ \pm $ 0.06 & 0.022 $ \pm $ 0.003 & 11.49 $ \pm $ 0.03 & -11.8 $ \pm $ 0.09 &  $<$ 9.9 \\
22719 & EGS & 1.58 & 1.8    $ \pm $ 0.5 & 0.40  $ \pm $ 0.09 & 0.012 $ \pm $ 0.003 & 10.96 $ \pm $ 0.08 & -11.7 $ \pm $ 7.4 &  $<$ 1.0 \\
7884 & COSMOS & 2.11 & 2.0  $ \pm $ 0.3 & 0.38  $ \pm $ 0.06 & 0.013 $ \pm $ 0.004 & 11.35 $ \pm $ 0.02 & -10.7 $ \pm $ 0.08 &  $<$ 9.6 \\
32707 & UDS & 1.65 & 2.1    $ \pm $ 0.3 & 0.31  $ \pm $ 0.08 & 0.014 $ \pm $ 0.004 & 11.20 $ \pm $ 0.04 & -11.5 $ \pm $ 0.10 &  $<$ 13.4 \\
17364 & COSMOS & 1.53 & 3.0 $ \pm $ 0.4 & 0.13  $ \pm $ 0.07 & 0.015 $ \pm $ 0.003 & 10.94 $ \pm $ 0.02 & -11.5 $ \pm $ 0.07 &  $<$ 1.7 \\
17089 & COSMOS & 1.53 & 3.4 $ \pm $ 0.2 & 0.30  $ \pm $ 0.06 & 0.013 $ \pm $ 0.002 & 11.45 $ \pm $ 0.02 & -12.0 $ \pm $ 0.07 &  $<$ 2.9 \\
\enddata
\tablecomments{More properties, including coordinates, morphological and dynamical measurements, are available in Table 2 of \citetalias{belli17mosfire}.}
\tablenotetext{a}{Derived from spectral fitting adopting a constant + tau model. The value and uncertainty are given by, respectively, the median and standard deviation of the posterior distribution.}
\tablenotetext{b}{Averaged over the past 100 Myr.}
\tablenotetext{c}{3-sigma upper limit on star formation rate, measured from the dust-corrected \Halpha\ flux.}
\tablenotetext{d}{Detected in the X-ray.}
\vspace{-5mm}
\end{deluxetable*}

We are able to obtain meaningful fits using each of the six SFH models for every galaxy in our sample except for one, ID 35616. Despite the fact that the MOSFIRE spectrum of this galaxy has a remarkably high signal-to-noise ratio, none of the six models of SFH is able to reproduce the photometry; for example adopting a tau model we obtain a reduced chi-square (for the photometry only) of 4.3, which is a clear outlier given that the rest of the sample has a mean chi-square of 1.1. Since the \HST\ imaging shows that this galaxy is clearly interacting with a close companion (see Figure 2 of \citetalias{belli17mosfire}), it is likely that some of the photometric measurements are contaminated. Furthermore, this object is also the only one in the sample with a strong \Halpha\ emission, as we discuss below in Section \ref{sec:emission_lines}. For these two reasons we exclude 35616 from the remainder of our analysis.

The 23 objects for which we obtain a good spectroscopic and photometric fit are shown in Figure \ref{fig:sample}. For each galaxy we plot the MOSFIRE spectrum with the best-fit template, and the median SFH we obtain when adopting a constant + tau model. The galaxies are sorted by their median age, from the youngest (top) to the oldest (bottom). The sample clearly spans a variety of stellar ages. A few galaxies, for which the fit indicates ages smaller than 1 Gyr, present strong high-order Balmer lines, indicative of populations dominated by B and A stars. The vast majority of the galaxies in the sample, however, show moderate Balmer absorption lines and a relatively strong 4000 \AA\ break, indicating the presence of both young and old stars. These galaxies can easily be sorted in age by looking at the relative strength of the two \CaII\ absorption lines \citep[see, e.g.,][]{almeida12}. If a significant population of stars younger than 1 Gyr is present, then the blend of the Ca H and H$\epsilon$ lines is deeper than Ca K. A further test of the relative ordering of these spectra comes from the H$\delta$ absorption line, which is stronger for younger populations. This qualitative assessment shows that the results of the fits are consistent with the rest-frame optical spectra, and are not dominated by the photometry.

In addition to the SFH, our spectral fits also determine the stellar mass, dust attenuation, and stellar metallicity, which are listed in Table~\ref{tab:sample} for the entire sample. The stellar metallicities span the range between half solar and slightly extra-solar, but their posterior distributions are broad, meaning that this quantity cannot be robustly measured by the spectral fits. We choose to include it as a free parameter anyway to avoid biasing the results for the other properties.

\subsection{The Effect of the SFH Model on the Results}
\label{sec:compare_sfh}

The example in Figure \ref{fig:spectrum_example} shows that the six SFH models can produce different results when applied to the same object, in terms of both the best-fit template and the recovered physical properties, such as the stellar age.
One of the goals of the present analysis is to compare different SFH models and to assess the extent to which the choice of the model affects the results of the spectral fit.

\begin{figure}[htbp]
\centering
\vspace{3mm}
\includegraphics[width=0.47\textwidth]{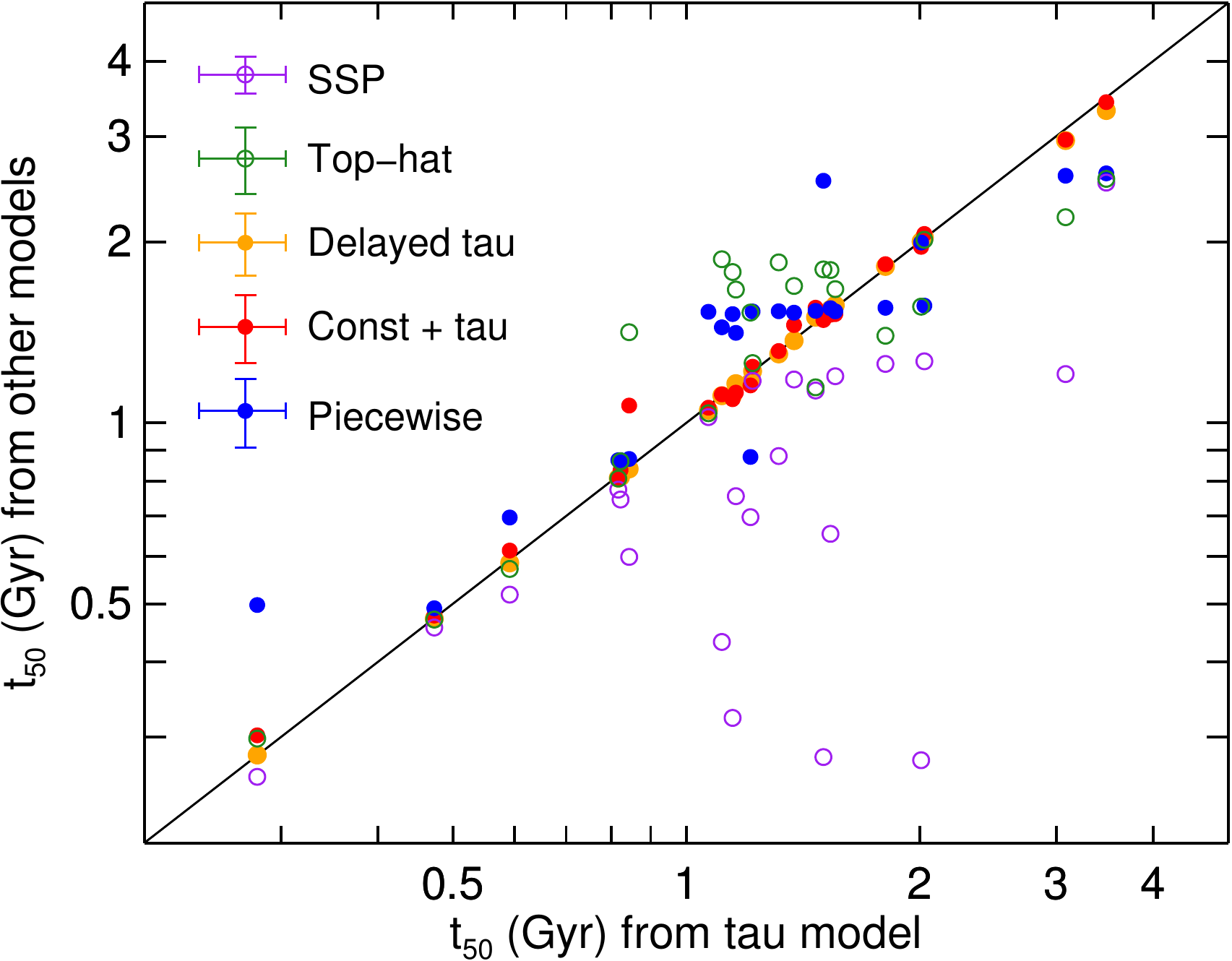}

\vspace{3mm}
\includegraphics[width=0.47\textwidth]{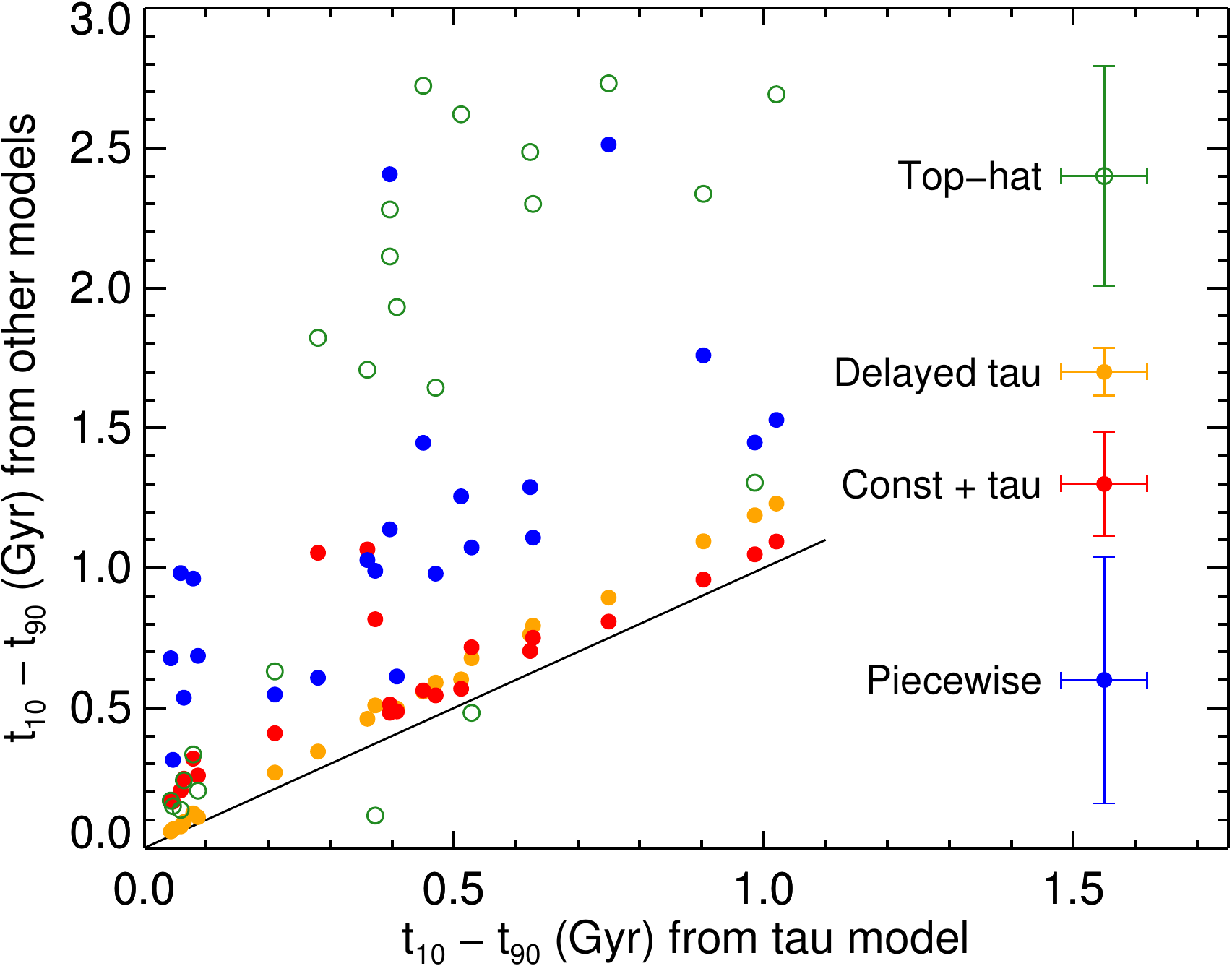}

\vspace{3mm}
\caption{Top: median stellar age $t_{50}$ measured using different SFH models, versus the value obtained adopting the tau model. Bottom: similar comparison for the quantity $t_{10}-t_{90}$, a proxy for the length of the star formation episode. This quantity cannot be measured for simple stellar populations. In both panels, each point corresponds to the median value of the posterior, and the median uncertainty for each SFH model is shown by the error bars in the legend.}
\label{fig:compare_sfh}
\end{figure}

Since each SFH model has a different number and type of parameters, we need to define non-parametric quantities in order to be able to compare the results among one another. Following \citet{pacifici16}, we define the characteristic time $t_X$ as the lookback time at which $X$\% of the stellar mass was already formed: $\int_{t_X}^{\infty} \mathrm{SFH}(t_\mathrm{lb}) dt_\mathrm{lb} = X/100$, where $t_\mathrm{lb}$ is the lookback time. We start by considering $t_{50}$, i.e., the median stellar age. We calculate the value of $t_{50}$ using the six SFH models, and show the results for the entire sample in the top panel of Figure \ref{fig:compare_sfh}. Since the tau model is the most widely used, we take it as a reference and plot, for each galaxy, the median ages obtained with the other models versus the median age obtained from the tau model. Clearly, the SSP model systematically underestimates the median age. This is a well known issue, due to the fact that the SSP model effectively returns the stellar population age weighted by light instead of mass. Interestingly, the other five models yield consistent results, showing that the galaxies in our sample span an order of magnitude in stellar ages, from 300 Myr to 3 Gyr. The delayed tau and the constant + tau models are in extremely good agreement with the tau model, while the top-hat and the piecewise models present a slightly larger scatter. The piecewise model shows an abundance of points around $t_{50} \sim 1.5$ Gyr, which is likely an artificial effect due to the positioning of the nodes (see Figure \ref{fig:spectrum_example}).

\begin{figure}
\centering
\includegraphics[width=0.40\textwidth]{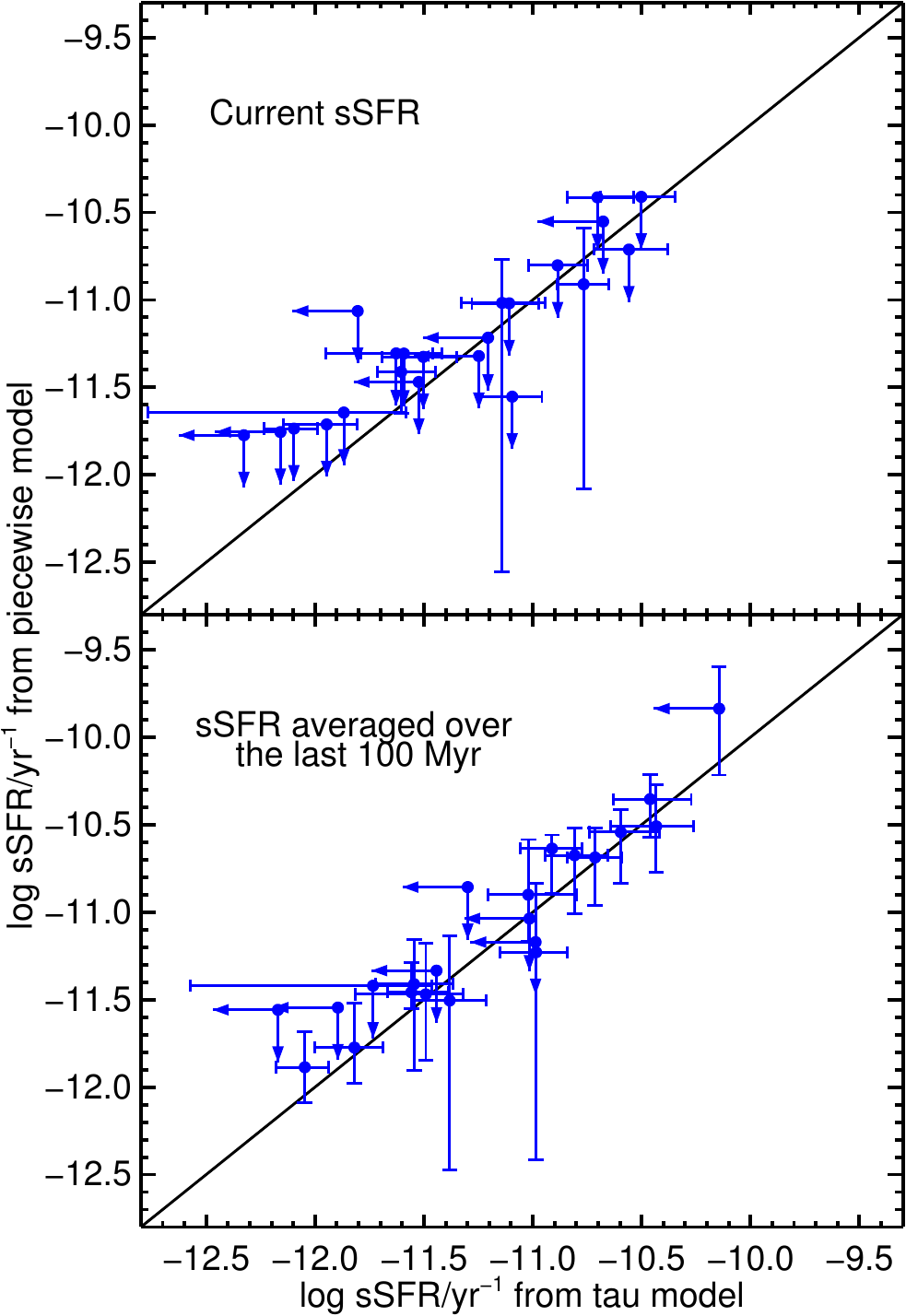}
\caption{Top: comparison of the current sSFR obtained using the piecewise model and using the tau model. Bottom: same comparison, but for the sSFR averaged over the past 100 Myr. Arrows indicate 95\% upper limits.}
\label{fig:compare_ssfr}
\end{figure}

Next, we consider the quantity $t_{10}-t_{90}$, which is the time it took to assemble the central 80\% of stellar mass of a galaxy, and can be considered as a proxy for the length of the star formation activity. The bottom panel of Figure~\ref{fig:compare_sfh} shows that different SFH models yield wildly different estimates for $t_{10}-t_{90}$. In particular, the top-hat produces very long bursts, often extending all the way to the limit set by the age of the universe. The tau model, instead, produces very short bursts, which is not surprising given its simple parameterization. The piecewise model yields intermediate results with large uncertainties, which are likely more realistic given its high degree of flexibility. The SSP yields zero by definition, and is not shown in the figure. We conclude that the length of the SFH is not as robust as the median age, and its measurement depends strongly on the assumed SFH model.

Another important property of quiescent galaxies is their current level of star formation activity. For each SFH model we divide the star formation rate (given by Equation~\ref{eq:SFR}) by the stellar mass, thus obtaining the specific star formation rate (sSFR). The SSP and the top-hat models by definition have zero current star formation rate. The delayed tau and constant + tau models are very similar to the tau model, therefore we only compare the piecewise and tau models in the top panel of Figure~\ref{fig:compare_ssfr}. For most of the galaxies we are only able to obtain upper limits, indicating that the spectra and the photometry are not able to constrain the current sSFR. But if we use the sSFR averaged over the previous 100 Myr, shown in the bottom panel, then we are able to obtain measurements with meaningful error bars for most of the galaxies. Moreover, the piecewise and the tau model are in excellent agreement, confirming that the spectra are indeed able to detect the level of star formation in the recent, but not immediate, past. The values of sSFR we obtain range between $10^{-12}$ and $10^{-10}$ yr$^{-1}$, corresponding to mass doubling times longer than three times the age of the universe at $z\sim2$, consistent with the quiescent nature of the sample.

The most robust measurements are therefore the median age $t_{50}$ and the sSFR averaged over the last 100 Myr, which are listed for each galaxy in Table~\ref{tab:sample}. When looking at parameters that are not directly related to the SFH, different models give consistent results. The stellar mass is particularly robust, with a difference between the piecewise and the tau model of only $0.02 \pm 0.04$ dex (respectively mean and standard deviation). Discrepancies in dust extinction ($-0.04 \pm 0.10$ mag) and stellar metallicity ($7 \pm 15$\%) are also small.

\begin{figure}
\centering
\includegraphics[width=0.45\textwidth]{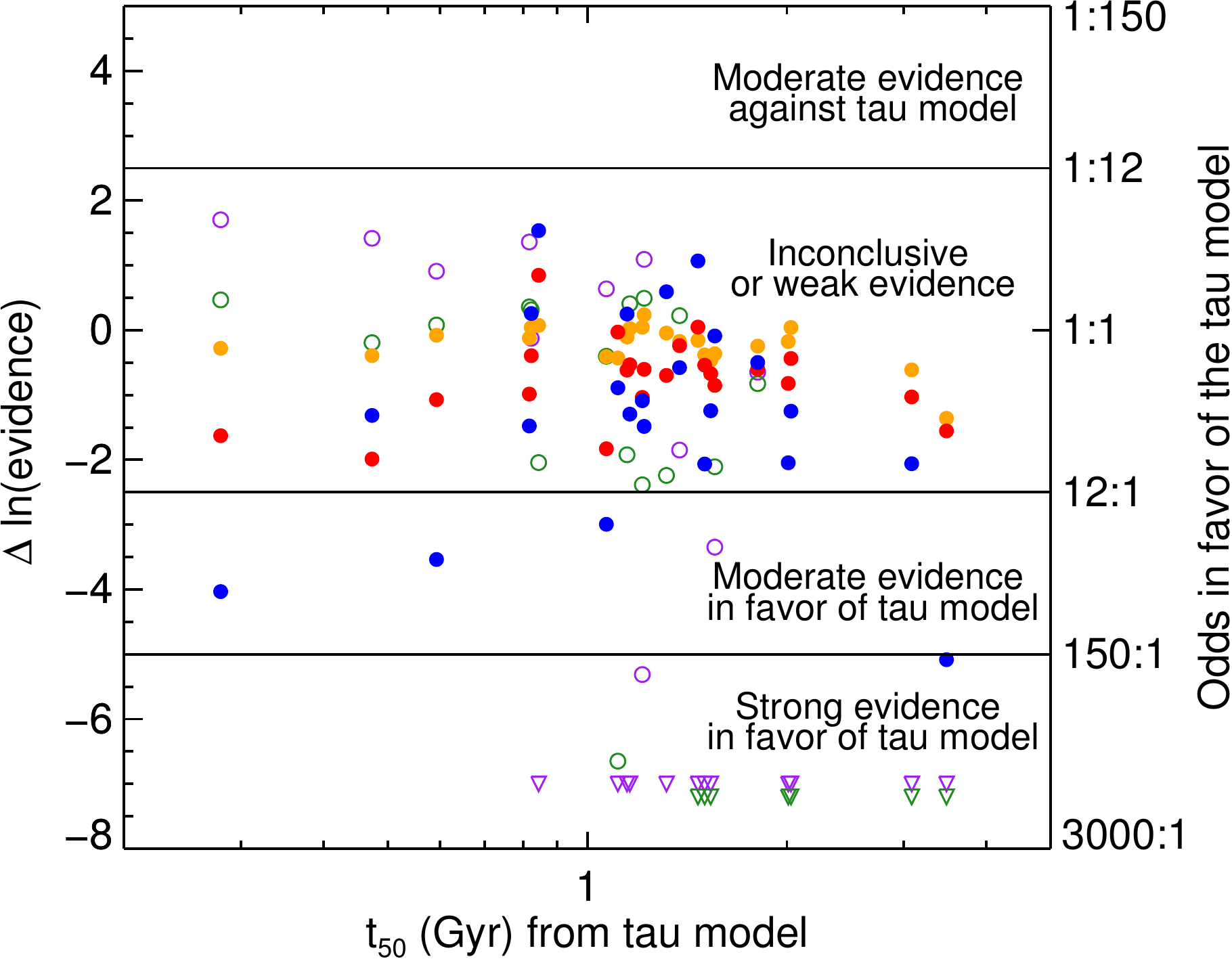}
\caption{Logarithmic difference between the evidence for a given model and the evidence for the tau model, as a function of median stellar age. Each color represents one of the five SFH models, as listed in Figure~\ref{fig:compare_sfh}. Triangles indicate upper limits. The right-hand $y$ axis gives the equivalent odds ratio, which is equal to $\exp(\Delta \ln \mathrm{evidence})$. Indicative regions of increasing evidence are marked, following \citet{trotta08}.}
\label{fig:compare_evidence}
\end{figure}

\subsection{Optimal SFH Models}

We now ask the question of whether the data favor any of the six models over the others. To answer this we take advantage of the nested sampling algorithm employed by \texttt{pyspecfit}, which in addition to the posterior distribution also returns the model \emph{evidence}. This quantity is the integral of the likelihood, weighted by the prior, over the full parameter space, and its value is proportional to the support of the data for a given model. In practice, model evidence is only relative: the ratio of the evidence of two models can be interpreted as the ratio of the odds in favor of one model against the other. Once again, we take the tau model as reference, and for each galaxy we define the quantity $\Delta \ln \mathrm{evidence}$ as the logarithmic difference between the evidence for a given model and the evidence for the tau model.

We show the distribution of $\Delta \ln \mathrm{evidence}$ for the entire sample, as a function of galaxy age, in Figure \ref{fig:compare_evidence}. This quantity can be directly translated to the odds that the tau model explains the data better than another model, and these are shown on the right-hand $y$ axis. We also mark the regions of the plot corresponding to \emph{weak}, \emph{moderate}, and \emph{strong} evidence according to a popular, though somewhat arbitrary, scale \citep[see, e.g.,][]{trotta08}. We find that the delayed tau and the constant + tau models fare as well as the tau model. On the other hand, the SSP and top-hat models are strongly disfavored. It is particularly interesting to note that the SSP actually fares slightly better than the tau model for young galaxies, but is unable to describe well older galaxies, for which a more extended SFH is clearly needed. Similar conclusions apply to the top-hat model.

The piecewise model is comparable to the tau model except at young ages, where it performs moderately worse, presumably because of its rather coarse binning. However, since the evidence is particularly sensitive to the adopted priors, the comparison of non-parametric and parametric models must be performed with care. For example, the maximum allowed stellar age for the piecewise model is fixed to 4 Gyr (see Section~\ref{sec:pyspecfit}); while for the parametric models we adopt the age of the universe at the observed redshift. To test whether this has any effect on the results, we run an additional spectral fit using a piecewise SFH where the last node, instead of being fixed at 4 Gyr, is equal to the age of the universe at the observed redshift. The resulting evidence is, on average, unchanged, and individual points move in Figure~\ref{fig:compare_evidence} by $\left| \Delta \ln \mathrm{evidence} \right| < 1.5$. We also test for the width of the priors, which we vary by one order of magnitude in each direction (i.e., adopting LogNormal(0,1) and LogNormal(0,3) on each of the SFH nodes, see Table~\ref{tab:SFH}). The quantity $\Delta \ln \mathrm{evidence}$ tend to decrease in both cases, on average by 1.0 and 0.5 respectively, and the result shown in Figure~\ref{fig:compare_evidence} is qualitatively unchanged. The physical properties, such as median ages and stellar masses, are remarkably stable. For an in-depth discussion of the effects of the prior when using non-parametric models, we refer to \citet{leja18nonparametric}.

Keeping in mind that the evidence is weighted by the size of the parameter space, we conclude that the extra degrees of freedom of the piecewise models (six free parameters as opposed to the two of the tau model) are not warranted, since the final results are not measurably better than those obtained with simpler models. This means that the secondary bursts recovered by the piecewise model in some cases, such as the one shown in Figure~\ref{fig:spectrum_example}, are possible, but are not required by the data. We caution, however, against a simplistic interpretation of this result: in the context of SFH models, having more degrees of freedom than required by the data is not necessarily a poor choice \citep[see, e.g.,][]{simha14}. For example, if one is interested in the measurement of a specific parameter, such as the median stellar age, it may be a good idea to allow more freedom than strictly required by the data.

Our analysis can be summarized as follows: the tau, delayed tau, and constant + tau models yield consistent results, whereas the SSP and top-hat models show clear limitations in reproducing the data. The piecewise model is the most flexible one, but does not describe the data significantly better than the simple tau model. It seems that the constant + tau model strikes a balance between simplicity and flexibility, and for this reason we choose it as our ``fiducial'' model.

\subsection{The H$\alpha$ and [NII] Emission Lines}
\label{sec:emission_lines}

\begin{figure}[tbp]
\centering
\includegraphics[width=0.40\textwidth]{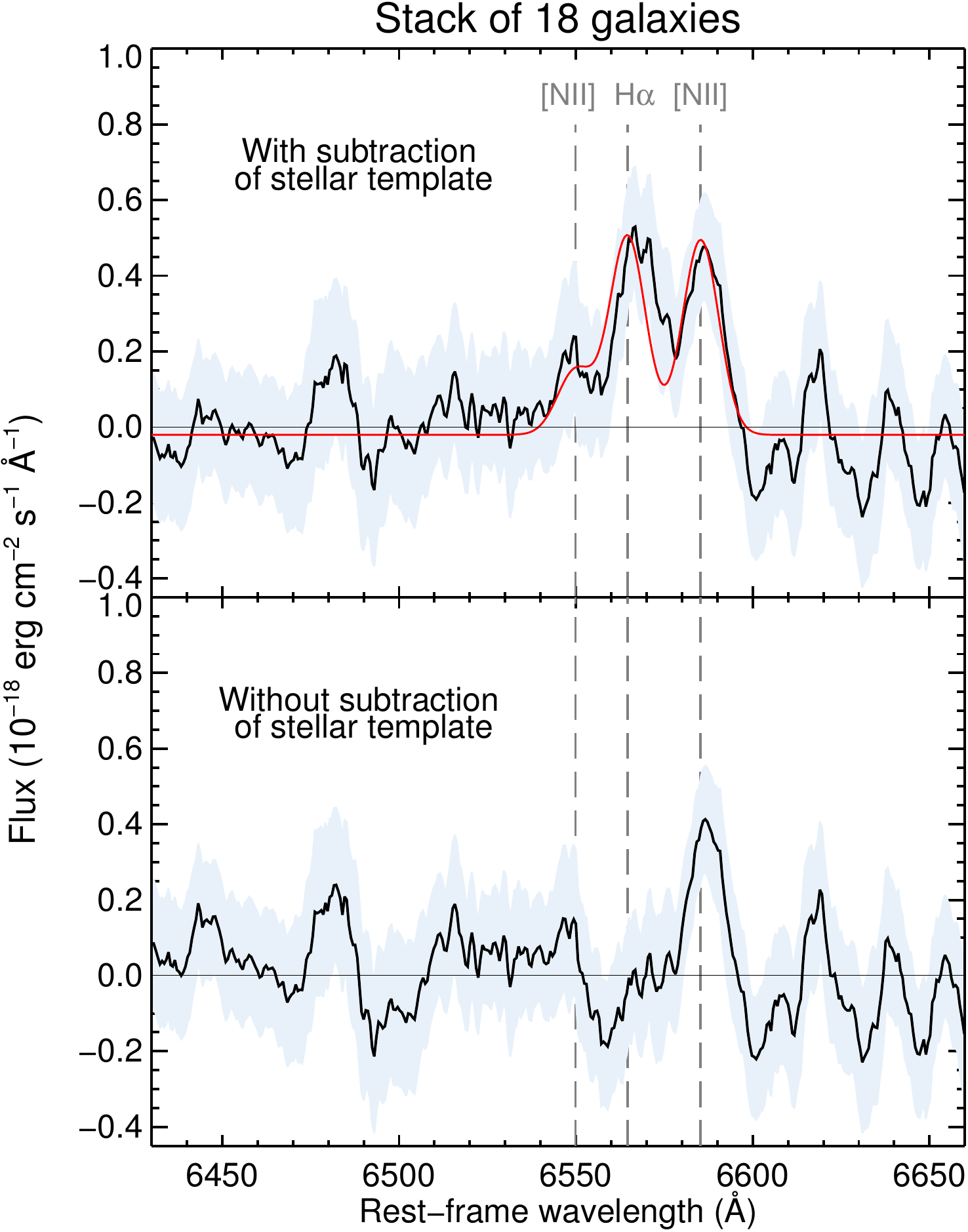}
\caption{Stacked \Halpha\ spectrum for 18 galaxies in our sample. The blue shaded region marks the uncertainty, and the red line is the Gaussian fit to the emission lines. The top panel shows the result when the best-fit stellar template is subtracted from each spectrum before stacking; the bottom panel shows the result obtained when simply subtracting the median continuum from each spectrum.}
\label{fig:stack_Ha}
\end{figure}

Our additional, shallower $H$ and $K$-band spectroscopic observations target the \Halpha\ emission for 19 out of 24 galaxies in the sample. We find only one clear detection of \Halpha, for 35616, the object we removed from the main analysis. The \NII\ line is also detected and, although \Halpha\ is contaminated by a strong sky line, we are able to derive a flux ratio \NII/H$\alpha \sim 0.8$, which is high and suggests a contribution from active galactic nucleus (AGN) activity \citep[e.g.,][]{kewley13}, consistent with the fact that this is one of the three galaxies in the sample that are detected in the X-ray. About 10\% of quiescent galaxies at high redshift harbor this type of emission \citep{belli17kmos3d}. Since 35616 is in a close interaction, it is possible that the AGN activity is triggered by gas that has been tidally stripped from the companion.

For the remaining galaxies in the sample we scale the observed spectrum so that the median level of the continuum around \Halpha\ matches the best-fit template from the spectral fit, to ensure a proper flux calibration. We then subtract the best-fit template to remove the contribution of the stellar absorption. We calculate upper limits on the non-detections by integrating the error spectrum in a window of 500 km s$^{-1}$ (corresponding to a line width $\sigma \sim 200$ km s$^{-1}$ ) centered on the \Halpha\ wavelength. After correcting the line fluxes for the extinction due to dust, adopting the attenuation $A_V$ obtained from the spectral fit, we convert to values of star formation rate (based on a Chabrier IMF) following \citet{kennicutt98}. The 3-sigma upper limits on the star formation rates are listed in Table \ref{tab:sample}. The upper limits range from 1 to 15 \Msun/yr; the highest values correspond to cases in which strong sky lines enter the \Halpha\ spectral window. For each galaxy, the limit on the star formation rate derived from \Halpha\ is higher than the value predicted by the spectral fit.

To obtain a deeper measurement, we shift each of the 18 spectra to the rest frame and stack them. The result is shown in the top panel of Figure~\ref{fig:stack_Ha}: both \Halpha\ and \NII\ are clearly detected. The bottom panel shows what we obtain when stacking the spectra without subtracting the stellar template. In this case the emission from \NII\ is detected while \Halpha\ is marginally detected in absorption. Without a detailed template of the stellar continuum it would not be possible to measure such small amount of \Halpha\ emission.

We model the emission lines \Halpha, \NII6548, and \NII6584 as three Gaussians. We fix the wavelengths to their theoretical values, the flux ratio of the two \NII\ lines to 3, and set the velocity dispersion of the three lines to be identical. The best-fit of this model to the continuum-subtracted stack is shown in red in Figure~\ref{fig:stack_Ha}. We obtain an \Halpha\ flux of $(6.7 \pm 0.8) \cdot 10^{-18}$ erg s$^{-1}$ cm$^{-2}$, a flux ratio \NII/H$\alpha = 1.0 \pm 0.1$, and a velocity dispersion of $233 \pm 17$ km/s. One galaxy, 30475, has a marginally detected \NII\ line (it is also one of the three galaxies of the sample detected in the X-ray). Removing it from the stack has a minimal impact on the results. Further splitting the sample into two random subsets does not change the results of the fit in a statistically significant way, confirming that the ionized emission is not due to any individual object. We also test the effect of the SFH model on the subtraction of the stellar template: adopting the best-fit piecewise model instead of the constant + tau model we obtain nearly identical emission line measurements.

The high \NII/\Halpha\ ratio we obtain for the stacked spectrum rules out star formation activity as the main source of emission lines. In the local universe, quiescent galaxies commonly feature weak emission lines due to the ionization from old stellar populations \citep{cidfernandes11}. However the maximum \Halpha\ equivalent width predicted for this mechanism is 3~\AA, while we measure $4.7 \pm 0.6$~\AA\ in our stack. Therefore we conclude that the most likely source for the ionized gas emission is low-level AGN activity. Ongoing star formation could still contribute to the observed flux, which therefore can be used to derive a strict upper limit to the star formation rate. After correcting the observed \Halpha\ flux for the average value of dust extinction in the sample ($A_V=0.43$), we obtain a star formation rate upper limit of $0.9 \pm 0.1$ \Msun/yr, which confirms the truly quiescent nature of the sample. Dividing by the average stellar mass we obtain $\log \mathrm{sSFR} \sim -11.1$, which is fully consistent with the values derived from the spectral fits and shown in Figure~\ref{fig:compare_ssfr}.


\begin{figure*}[htbp]
\centering
\vspace{5mm}
\includegraphics[width=0.75\textwidth]{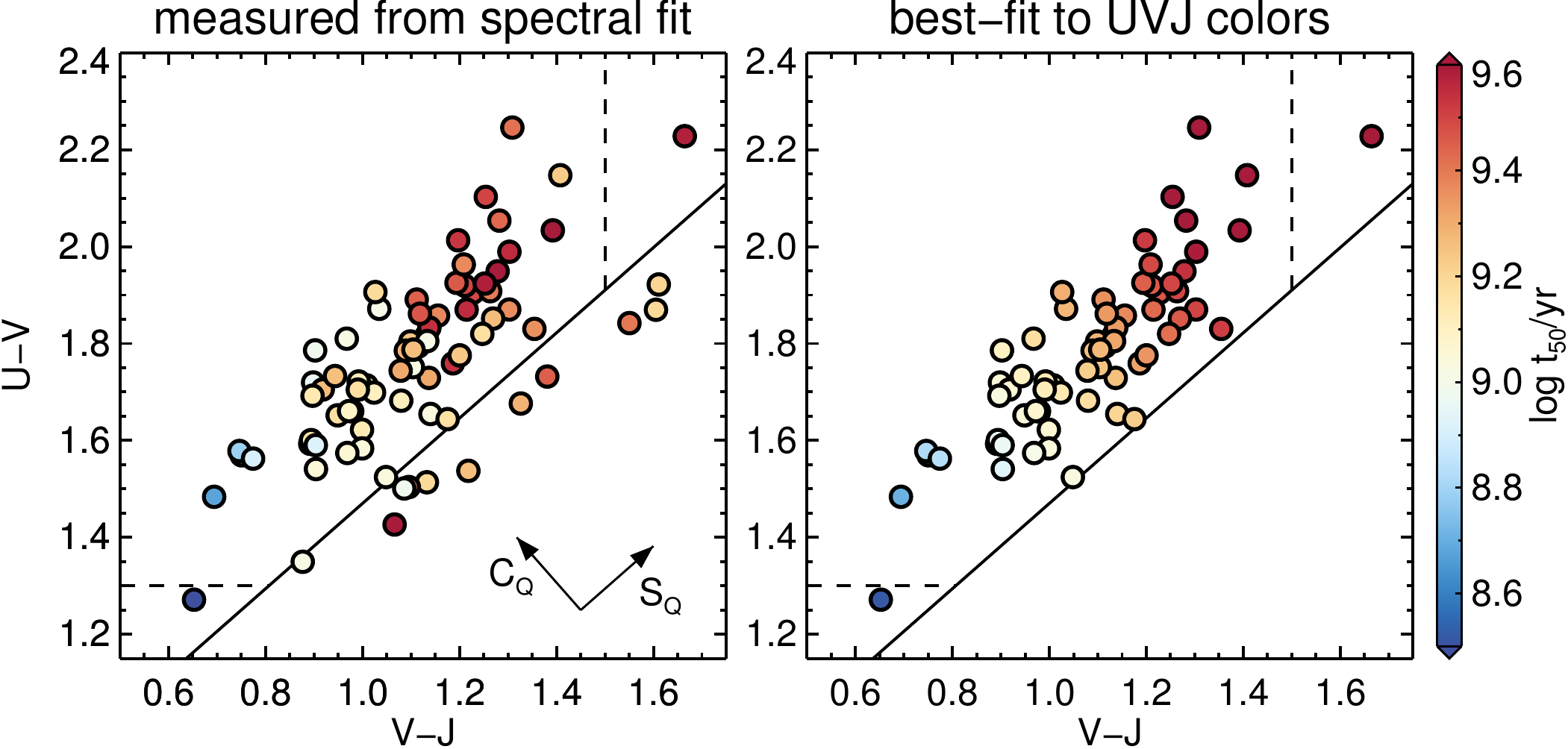}
\caption{Left: $UVJ$ diagram for both the MOSFIRE and LRIS samples, color-coded by the median stellar age $t_{50}$ as measured from fitting models to the spectroscopic and photometric data. The rotated coordinates \Sq\ and \Cq\ are shown in the bottom right corner. Right: same, but color-coded by the median age inferred from the $UVJ$ colors using Equation~\ref{eq:inferred_age}. In this panel only galaxies within the quiescent region, defined by the diagonal black line, are shown. The dashed lines mark the edges of the quiescent population proposed by \citet{muzzin13}, which we do not use in our analysis.}
\label{fig:ages_UVJ}
\end{figure*}

\section{The Age--Color Relation}
\label{sec:age-color}

In this section we will develop a simple technique to estimate stellar ages for galaxies without using spectroscopic data, by taking advantage of the relatively tight relation between median age and rest-frame colors. This will enable the study of a much larger, photometric sample, which will be presented in the next section.

In order to probe a larger redshift range and to increase the sample size, for the present analysis we combine the MOSFIRE data with the LRIS observations at $1 < z < 1.5$ from our previous studies \citep{newman10, belli14lris, belli15}. We perform the spectroscopic fits to the LRIS sample as described in Section \ref{sec:fit}, adopting a constant + tau model. We remove objects with low signal-to-noise or poor fits, and also objects that are outside the CANDELS field of view and therefore are not present in the 3D-HST catalog; this leaves 56 galaxies. The results of the spectral fitting for these objects are given in Appendix~\ref{sec:appendix_lris}. Combining the two Keck samples we obtain a homogeneous data set of 79 galaxies spanning the redshift range $1 < z < 2.5$ and the stellar mass range $10.5 < \log \Mstar/\Msun < 11.5$.

In the left panel of Figure~\ref{fig:ages_UVJ} we show the distribution of the median stellar ages $t_{50}$ for the combined Keck sample on the $UVJ$ diagram, i.e., the rest-frame $U-V$ vs $V-J$ plane. This is a useful diagram in which quiescent and star-forming systems tend to form two distinct sequences \citep[e.g.,][]{wuyts07}. We take the rest-frame colors from the 3D-HST catalog, which are therefore independent of our Keck spectra. The figure shows that the distribution of stellar ages is remarkably regular, and that the red sequence is a sequence in age.
This is not a new result: we already showed this using our LRIS sample at $1 < z < 1.5$ in \citet{belli15}, and an independent study by \citet{dominguezsanchez16} confirmed this finding. At $z\sim2$, \citet{whitaker13} and \citet{mendel15} used stacked spectra to show that the blue end of the sequence consists of younger galaxies. However, our sample presents two important advantages over previous studies: it has a relatively large number of individual measurements, so that we do not have to rely on stacking; and it is based on a robust estimate of the stellar age, that does not depend on the SFH assumptions, as we have shown in Section~\ref{sec:fit}. We point out that we also detect the age trend along the red sequence in the MOSFIRE sample alone, as shown in Appendix~\ref{sec:appendix_uvj}.

The diagonal black line marks our definition of quiescent region, which we take from \citet{muzzin13}. The original definition included a vertical boundary at $V-J=1.5$ and a horizontal one at $U-V=1.3$, shown as dashed lines. Recent studies have suggested that these additional constraints exclude galaxies that are quiescent and dusty \citep[e.g.,][]{whitaker15}, and galaxies that have been quenched abruptly \citep{merlin18}, respectively. Interestingly, the age trend does seem to continue beyond the very blue and very red ends of the quiescent box. For these reasons we do not consider these edges and simply use the diagonal line to define the quiescent population.
The smooth age trend breaks down just outside this diagonal line. This happens for a variety of reasons, including a higher dust attenuation and a stronger contribution of the \OII\ emission line to the $U$-band flux of star-forming galaxies. The latter effect depends strongly on the templates used to derive the rest-frame colors: when adopting purely stellar templates, as we did in \citet{belli15}, these objects fall inside the quiescent box. For consistency, throughout this work we use rest-frame colors derived fitting exclusively the broadband photometry and using the same templates \citep[from][]{brammer08}.

\begin{figure}[tbp]
\centering
\includegraphics[width=0.45\textwidth]{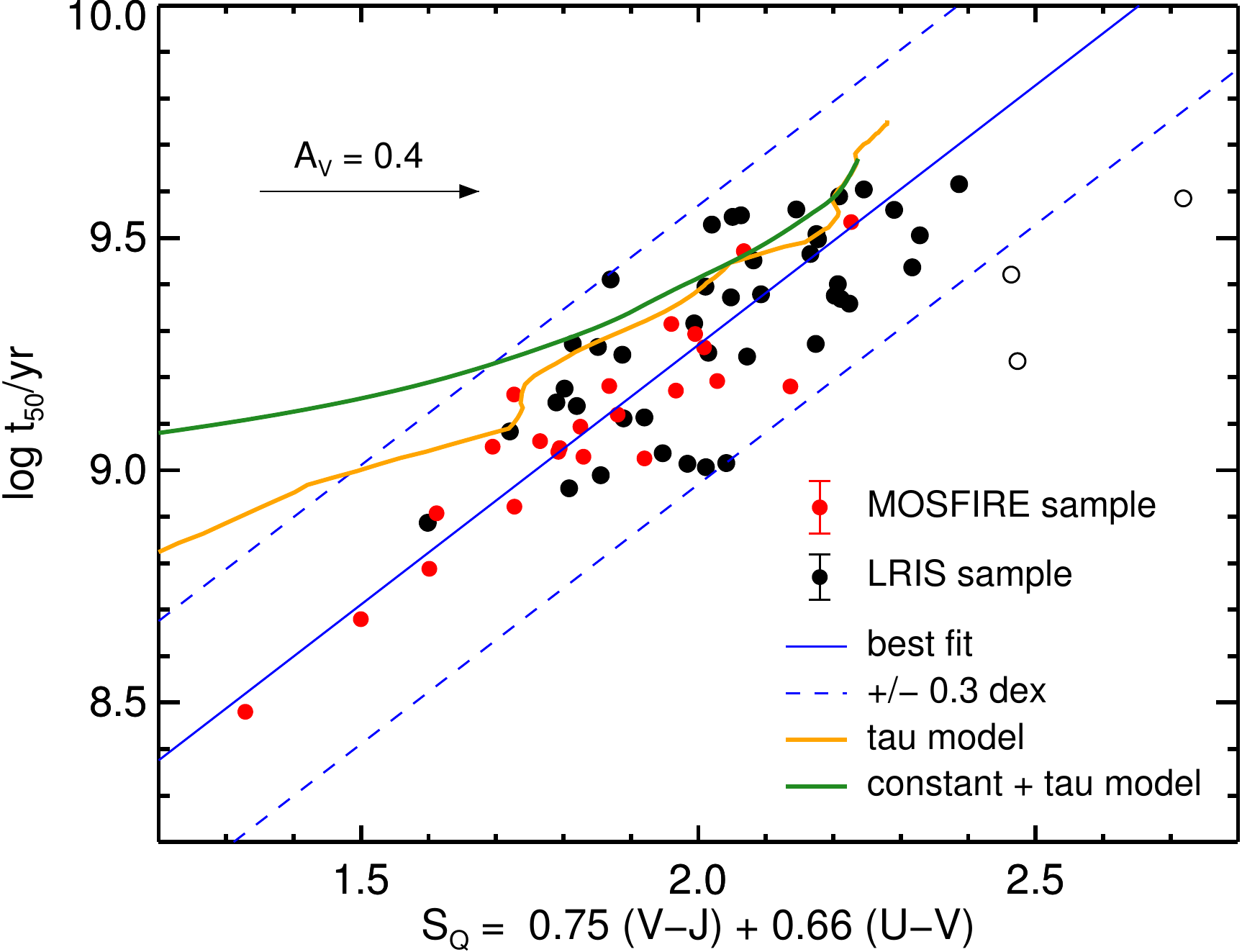}
\caption{Measured median stellar age as a function of the \Sq\ coordinate, for both the MOSFIRE (red) and LRIS (black) samples. The median age uncertainties are shown by the error bars in the legend; empty points are outliers and are excluded from the fit. The blue line shows the best-fit given in Equation~\ref{eq:inferred_age}, and the dashed blue lines mark the region within 0.3 dex from the best-fit. The tracks for two simple models of star formation histories are also shown: a tau model with $\tau=100$ Myr (orange) and a constant star formation of 2 Gyr followed by an exponential decline with $\tau=100$ Myr (green). The models stop 5.7 Gyr (the age of the universe at $z\sim1$) after the onset of star formation. The horizontal arrow shows the change in \Sq\ due to a dust attenuation $A_V=0.4$, which is the mean value measured for the galaxies in the Keck sample.}
\label{fig:inferred_age}
\end{figure}

To simplify the quantitative analysis of the age trend, we introduce a rotated system of coordinates on the $UVJ$ diagram, following \citet{fang18}. The rotated axes \Sq\ and \Cq, illustrated in the left panel of Figure~\ref{fig:ages_UVJ}, are respectively parallel and perpendicular to the boundary of the quiescent box as defined in \citet{muzzin13}, which has a slope of 0.88. Thus, the new coordinate system corresponds to a rotation of $\theta_\mathrm{Q} = \arctan 0.88 = 41\fdg 3 $, and is described by the following transformations:
\begin{equation}
  \label{eq:coords}
\begin{split}
  \Sq & = 0.75 \, (V-J) + 0.66 \, (U-V) \; , \\
  \Cq & = -0.66 \, (V-J) + 0.75 \, (U-V) \; .
\end{split}
\end{equation}
The angle $\theta_\mathrm{Q}$ is slightly different from that used by \citet{fang18}, $\theta=34\fdg 8$, which is appropriate for the sequence of star-forming galaxies. In order to distinguish the two sets of coordinate, we add the subscript ``Q'' to ours to indicate that these apply to the quiescent population. It is easy to see that the stellar ages are only a function of \Sq, and do not depend on \Cq. Selecting only the objects in the quiescent region, we show the relation between the logarithmic stellar age and rotated coordinate \Sq\ in Figure~\ref{fig:inferred_age}. Most galaxies lie on a remarkably tight relation. To better understand this relation, we also show the theoretical tracks for two simple, dust-free star formation histories: one is a tau model and the other is a model where the star formation rate is constant for 2 Gyr and then falls off exponentially. Both models are significantly offset from the observed distribution, particularly for ages younger than 1 Gyr. This is not surprising, since dust extinction has a strong impact on the value of \Sq, and our spectroscopic fits reveal the presence of some dust in most quiescent galaxies, with a mean value of $\left< A_V \right> = 0.4$, which corresponds to a reddening effect of $\left< \Delta \Sq \right> = 0.33$, shown by the horizontal arrow in the figure. We conclude that the observed tight relation between stellar age and rest-frame colors cannot be predicted by a simple model, and is actually due to a combination of stellar population aging and evolution of the dust content.

\begin{figure}[tbp]
\centering
\includegraphics[width=0.45\textwidth]{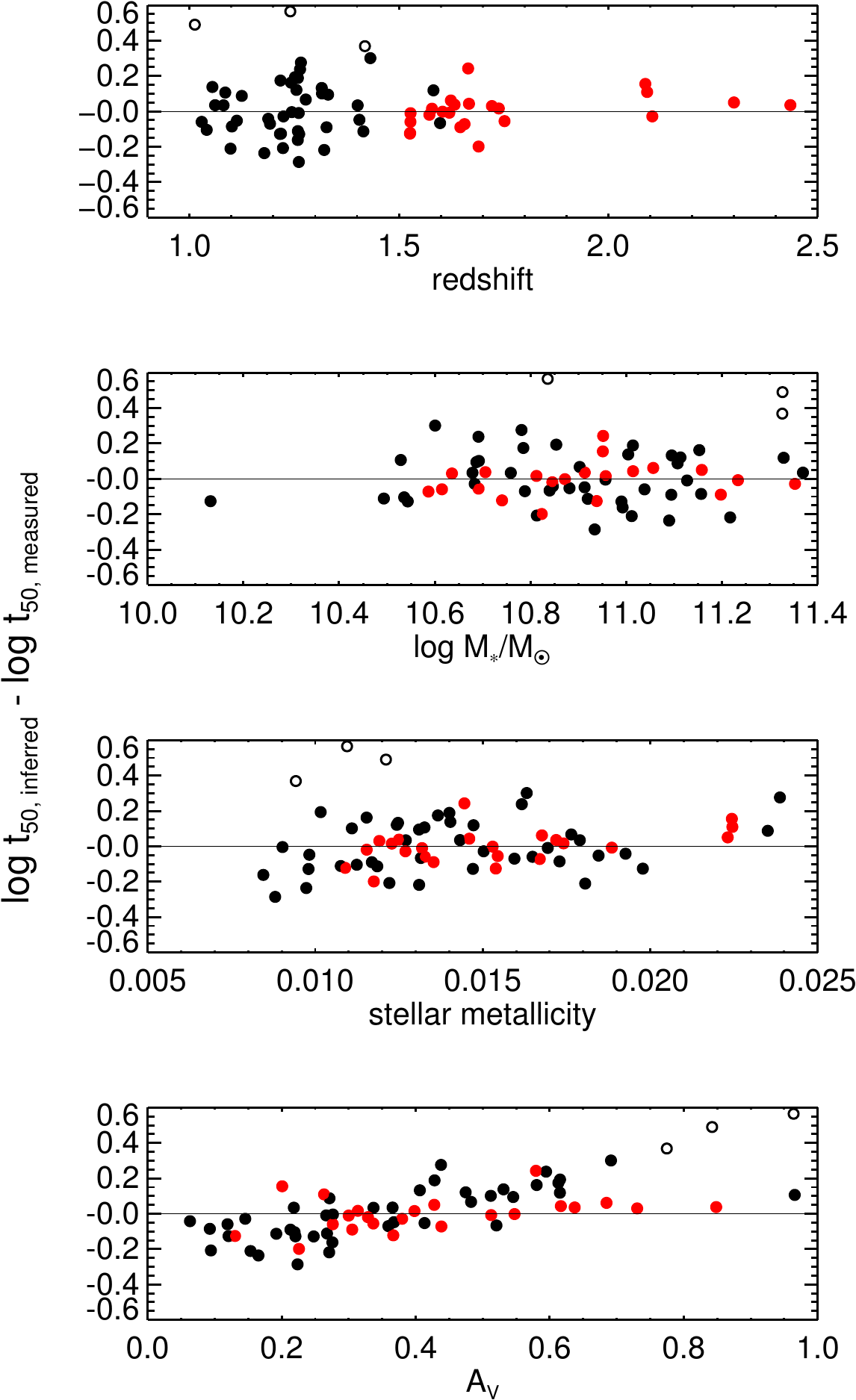}
\caption{Logarithmic difference between inferred and measured median ages, as a function of redshift, stellar mass, stellar metallicity, and dust attenuation $A_V$, from top to bottom. All physical properties are derived from the spectroscopic fit. Symbols as in Figure \ref{fig:inferred_age}.}
\label{fig:age_residuals}
\end{figure}

The model tracks reach a maximum value of \Sq\ around 2.3. The observed galaxies that have colors much redder than this must be affected by substantial dust extinction. This is particularly true for the three objects (shown as empty points) that are redder than $\Sq\sim2.4$, which are clearly outliers, and have dust attenuation values among the highest in the sample, $A_V \sim 0.8 - 1$. Excluding these objects, we fit a linear relation to the data, obtaining
\begin{equation}
\begin{split}
  \label{eq:inferred_age}
  \log \, t_{50}/\mathrm{yr} & = 7.03 + 1.12 \cdot \Sq \\
  & = 7.03 + 0.84 \, (V-J) + 0.74 \, (U-V) \; ,
\end{split}
\end{equation}
which is shown in blue in the figure. Virtually all galaxies lie within $\pm 0.3$ dex of this relation; the only exceptions are the three dusty galaxies. It is therefore reasonable to limit the validity of Equation~\ref{eq:inferred_age} to the range $\Sq < 2.4$. All galaxies redder than this value are particularly dusty, and a proper estimate of their stellar ages must rely on direct spectroscopic observations. Excluding the outliers, the standard deviation of the age residuals is 0.13 dex, or 35\%. This means that it is possible, in most cases, to predict the median stellar age of a quiescent galaxy at high redshift just by measuring the rest-frame $U-V$ and $V-J$ colors. The scatter of this relation is significantly larger than the typical age uncertainty (0.05 dex), meaning that spectroscopic data are still needed for a precise measurement. Nonetheless, Equation~\ref{eq:inferred_age} is a very effective tool for estimating stellar ages from photometric observations. We show this in the right panel of Figure~\ref{fig:ages_UVJ}, where we color code the Keck sample by the stellar ages inferred from the $UVJ$ colors using the best-fit relation, finding a striking agreement with the values measured from the spectroscopy, shown in the left panel.

To check that this age--color relation holds for the full population of quiescent galaxies, we plot the age residuals as a function of redshift, stellar mass, stellar metallicity, and dust attenuation in Figure~\ref{fig:age_residuals}. The residuals are remarkably flat in redshift and stellar mass, and show a small trend with metallicity and dust attenuation. This is expected, since both metallicity and dust can redden the observed SED of a galaxy, but the intensity of this effect is small, pointing once more at a correlation between different galaxy properties that has the effect of reducing the scatter in color for a given stellar age. However, given the poor constraint on the stellar metallicity obtained with our spectral fits, it is possible that part of the remaining scatter in the age-color relation is driven by metallicity.

In principle, from the last panel of Figure~\ref{fig:age_residuals} it is possible to derive a correction term that depends on the dust attenuation $A_V$ and that would lead to an even tighter relation between color and age. However, the physical properties shown in the figure are derived from the fit to both spectroscopy and photometry; in the absence of spectra the degeneracy between dust and age would lead to a substantially larger uncertainty on the dust attenuation \citep[see, e.g.,][]{belli15}, thus limiting the accuracy of such a correction.


\begin{figure*}[htbp]
\centering
\includegraphics[width=\textwidth]{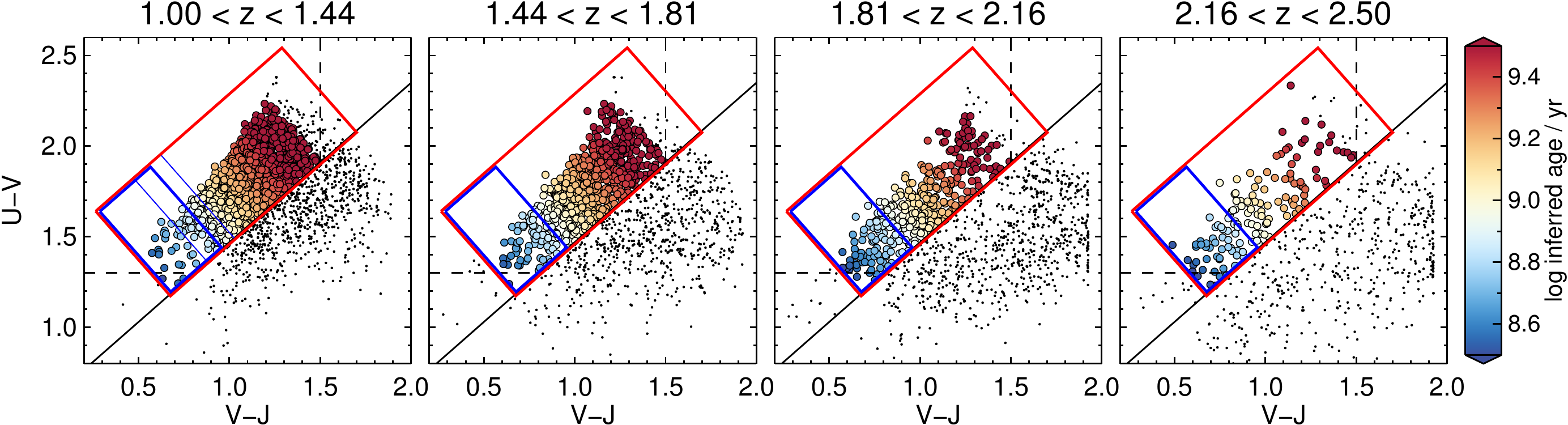}
\caption{Distribution of the rest-frame $UVJ$ colors for all galaxies with $\log \Mstar/\Msun > 10.8$ in the UltraVISTA survey, split into four redshift bins spanning equal comoving volume. The points are color-coded by inferred median ages using Equation~\ref{eq:inferred_age}. Black dots represent galaxies for which the inferred age cannot be calculated, either because they are outside the range in which the age--color relation has been calibrated, or because they are star-forming objects. The red box indicates the selection region for quiescent galaxies; the blue box that for post-starburst galaxies, defined as having 300 Myr $ < t_{50} < $ 800 Myr. In the first panel, thin blue lines show how the edge for the selection box changes when adopting a maximum age for post-starburst galaxies of 600 Myr or 1 Gyr.}
\label{fig:UVJ_uvista}
\end{figure*}

\begin{deluxetable*}{lcccc}
  \vspace{3mm}
\tabletypesize{\footnotesize}
\tablewidth{0pc}
\tablecaption{Galaxy Number Densities in the UltraVISTA Survey\label{tab:numden}}
\tablehead{
\colhead{} & \colhead{$1.00 < z < 1.44$} & \colhead{$1.44 < z < 1.81$} & \colhead{$1.81 < z < 2.16$} & \colhead{$2.16 < z < 2.50$}
}
\startdata
All Quiescent Galaxies & -3.69 $\pm$ 0.04 & -3.98 $\pm$ 0.05 & -4.27 $\pm$ 0.06 & -4.63 $\pm$ 0.07 \\
300 Myr $ < \log t_{50} < $ 1 Gyr & -4.79 $\pm$ 0.06 & -4.77 $\pm$ 0.06 & -4.71 $\pm$ 0.07 & -4.96 $\pm$ 0.08 \\
300 Myr $ < \log t_{50} < $ 800 Myr & -5.05 $\pm$ 0.07 & -4.96 $\pm$ 0.07 & -4.87 $\pm$ 0.07 & -5.10 $\pm$ 0.09 \\
300 Myr $ < \log t_{50} < $ 600 Myr & -5.41 $\pm$ 0.09 & -5.32 $\pm$ 0.09 & -5.13 $\pm$ 0.08 & -5.30 $\pm$ 0.10 \\
\enddata
\tablecomments{The quantities listed correspond to $\log N/\mathrm{Mpc}^{-3}$, where $N$ is the number density. Only galaxies with $\log \Mstar/\Msun > 10.8$ are considered.}
\end{deluxetable*}

\section{Number Density Evolution of the Quiescent Population}
\label{sec:numdens}

Now that we have a method to estimate the stellar ages without the need for spectroscopic data, we can explore the evolution of quiescent galaxies in large, photometric surveys. The goal of this section is to measure number densities and study their evolution with cosmic time, a task currently not practical with spectroscopic samples due to their relatively small size and complex selection function.

We consider the galaxy population in the UltraVISTA survey \citep{mccracken12}, which covers a wide area (1.62 deg$^2$) and thus enables the measurement of number densities with a high degree of precision. We take the \citet{muzzin13catalog} catalog of sources in the UltraVISTA field, and follow the sample selection performed by \citet{muzzin13} for studying the galaxy mass function: we exclude all objects that are fainter than $K_S=23.4$, flagged as stars, located in bad regions, or with contaminated photometry.
We also restrict our analysis to the redshift range in which the age--color relation has been calibrated, $1 < z < 2.5$, and to galaxies with $\log \Mstar/\Msun > 10.8$, which is the 95\% mass-completeness threshold at $z\sim2.5$ \citep{muzzin13}. The resulting sample consists of 5335 galaxies, and its distribution on the $UVJ$ diagram is shown in Figure~\ref{fig:UVJ_uvista}, split into four redshift bins. Galaxies are color coded according to their inferred median age, which we derive using Equation \ref{eq:inferred_age}. Star-forming galaxies, or quiescent galaxies outside the calibrated color range, are shown as black dots.
The four redshift bins have been defined so that each bin spans the same comoving volume. This means that the number of galaxies shown in the figure can be directly interpreted as a number density. For example, it can be easily seen that the population of massive quiescent galaxies grows substantially between $z \sim 2.5$ and $z\sim1$.

Since the quiescent population forms a sequence of increasing stellar age, the simplest model to explain the observations is one where all galaxies appear at the blue end and, as they age, move along the sequence towards the red end. This simple hypothesis has strong implications for galaxy quenching, as we will discuss below. It is therefore important to test whether the data are in agreement with this model. Combining number densities and stellar ages, we are now able to perform this test.

We start by selecting the population of quiescent galaxies, using the red box in Figure~\ref{fig:UVJ_uvista}. Then we select, among the quiescent galaxies, those that have young stellar ages. Following popular terminology, we refer to these as \emph{post-starburst} galaxies, since they are observed a short time after their main phase of star formation (independently of whether that was a burst). Looking at the age--color relation for the Keck sample, shown in Figure~\ref{fig:inferred_age}, we find that an age threshold of 800 Myr gives a pure selection, meaning that virtually all galaxies with an inferred age below this threshold happen to have a spectroscopic age that is also below this value. We also limit the definition to galaxies that are older than 300 Myr, the youngest age we measure in the Keck sample, simply because the age--color relation has not been tested beyond this limit. The selection for post-starburst galaxies is therefore 300 Myr $ < t_{50} < $ 800 Myr, and is shown as a blue box in Figure~\ref{fig:UVJ_uvista}.

\begin{figure}[tbp]
\vspace{5mm}
\centering
\includegraphics[width=0.45\textwidth]{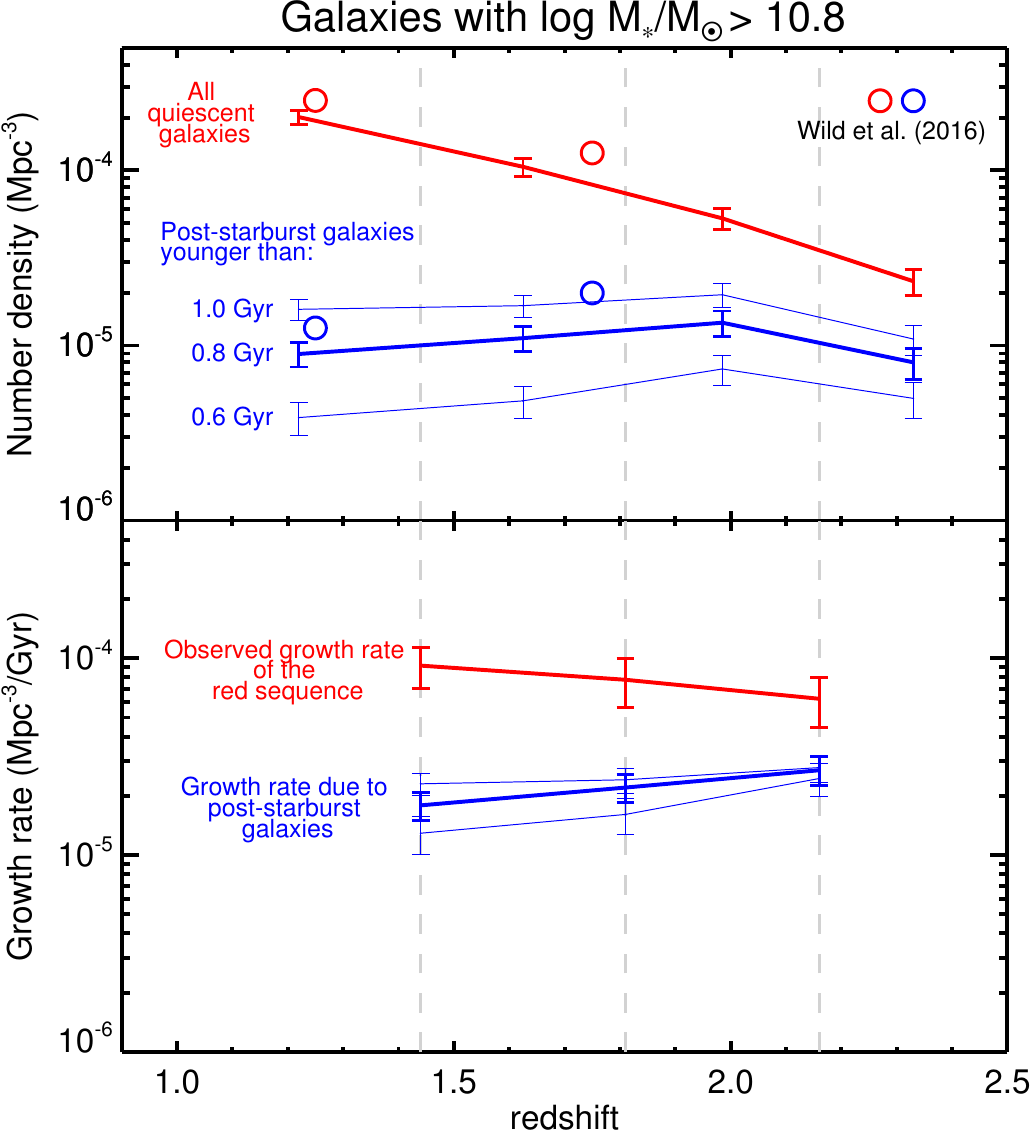}
\caption{Top: number density of quiescent galaxies (red) and post-starburst galaxies (blue) with $\log \Mstar/\Msun > 10.8$ in the UltraVISTA survey. Circles mark the values reported by \citet{wild16}. Bottom: observed growth rate of the quiescent population (red) and rate of growth due to the flux of post-starburst galaxies (blue). In both panels, the thin blue lines correspond to changes in the threshold age used to select post-starburst galaxies. The vertical dashed lines mark the redshift bins, which are the same as those in Figure~\ref{fig:UVJ_uvista}.}
\label{fig:uvista_numden}
\end{figure}

We count the number of objects in each of the selection regions, and divide by the comoving volume probed in each redshift bin, obtaining number density measurements. These are listed in Table~\ref{tab:numden} and shown in the top panel of Figure~\ref{fig:uvista_numden} as a function of redshift, for the quiescent galaxies (red line) and the subset of post-starburst galaxies (thick blue line). Uncertainties are calculated by combining in quadrature the Poisson error with the cosmic variance, which is calculated following \citet{moster11}. Since the value of 800 Myr is arbitrary, we vary the post-starburst selection by moving the threshold to 1 Gyr and 600 Myr. The corresponding number densities are plotted as thin blue lines, and the effect of this change on the edge of the selection box is shown in the first panel of Figure~\ref{fig:UVJ_uvista}. We find that post-starburst galaxies have a roughly constant number density over the entire redshift range. However, as the overall population of quiescent galaxies grows rapidly with cosmic time, the subset of post-starburst galaxies becomes increasingly less relevant, going from being 34\% of the total quiescent population at $z\sim2.5$ to just 4\% at $z\sim1$.

Next, we measure the rate of growth of the red sequence, by taking the difference of the number density of quiescent galaxies in adjacent bins, divided by the cosmic time interval between the centers of the bins. This is plotted in red in the bottom panel of Figure~\ref{fig:uvista_numden}. We then compare this to the flux of galaxies going through the post-starburst region. This is given by the number density of post-starburst galaxies divided by the time it takes them to cross the post-starburst selection region. Assuming a passive evolution, galaxies will take 500 Myr to cross the post-starburst region, going from an age of 300 Myr to an age of 800 Myr. We show this number density growth with a thick blue line in the figure.
It is clear that the flux of galaxies through the post-starburst region is not sufficient to explain the observed growth of the quiescent population. At $z \sim 2.2$ the observed growth rate is about twice that inferred from the post-starburst population. At $z\sim1.4$ this ratio has increased to $\sim 5$ times.
Moreover, this result does not depend on the age threshold used in the definition of the post-starburst population, as shown by the thin blue lines in Figure~\ref{fig:uvista_numden}. Changing the age threshold has a large impact on the measured number densities, but not on the growth rates, because of the corresponding change in the time spent in the post-starburst region. This is an indirect validation of our inferred age technique.

Since the flux of post-starburst galaxies is much smaller than the observed growth of the quiescent population, we must conclude that many galaxies appear on the red sequence when their median ages are already older than 1 Gyr. On the $UVJ$ diagram, this would correspond to crossing the diagonal line dividing quiescent and star-forming systems, without ever entering the post-starburst region. This is the main result of our study, and has important consequences for the study of galaxy quenching. Since post-starburst galaxies are both quiescent and young, they must have been quenched in a very rapid way, on timescales of the order of a few hundred Myr. However, our result shows that most massive galaxies did not undergo this phase, and instead must have been quenched slowly, to allow their stellar populations to age while they were still star forming.

The growth rates that we measured are based on ages inferred from photometry alone, which cannot be as accurate as spectroscopic measurements. However, given the large galaxy population used in the derivation, the precision in the inferred age of a single object is not relevant. What matters is that we are not missing post-starbust galaxies that are shifted outside the selection box by heavy dust reddening. Figure~\ref{fig:inferred_age} shows that no such galaxies are present in the Keck sample. The outliers in the age--color relation are dusty but also old, and are so rare (see black points in Figure~\ref{fig:uvista_numden}) that their impact on the number density calculation is limited. On the other extreme of the color distribution, some galaxies in the UltraVISTA survey are bluer than the youngest objects in the Keck sample, and have not been considered in the number density calculation. This is also a rare population, and extending the selection region to include these objects does not change the results significantly.
We also assumed, in our derivation, that galaxies evolve passively during the post-starburst phase. We note that this is a conservative assumption: any deviation from passive evolution, such as frosting or rejuvenation, will make the colors bluer, increase the visibility time for the post-starburst galaxies, and therefore decrease their contribution to the growth of the red sequence, strengthening our result.

In our analysis we neglected the contribution of mergers to the number density evolution of quiescent galaxies. This is a reasonable approximation, given that in the time spanned by our redshift bins only 1 in 10 galaxies will undergo a major merger \citep[e.g.,][]{man16merging}. In Appendix~\ref{sec:appendix_merging} we carry out a quantitative assessment of merging, which causes two distinct and opposite effects. On the one hand, massive quiescent galaxies merge together thus decreasing the number density of the quiescent population; on the other hand, quiescent systems with $\log \Mstar/\Msun < 10.8$ can grow in mass entirely via gas-poor mergers, therefore increasing the number density of the population considered in our analysis. In Appendix~\ref{sec:appendix_merging} we show that these two effects are one to two orders of magnitude smaller than the observed growth of the quiescent population over $1<z<2.5$, and can therefore be neglected.

Our result differs from that of a previous study by \citet{whitaker12}, who find that the growth of the red sequence at $z>1$ is consistent with the aging of post-starburst galaxies. However, Whitaker et al. base their selection of post-starburst galaxies on the results of SED fitting, obtaining a window on the $UVJ$ diagram that is substantially wider than ours.
In another study of post-starburst galaxies, \citet{wild16} adopt a selection based on a principal component analysis of galaxy SEDs. While this selection is in principle completely independent from our method, it gives very similar results in terms of distribution of $UVJ$ colors \citep{wild14}. Not surprisingly, we find number densities for both the quiescent and the post-starburst populations in agreement with those measured by \citet{wild16}, shown as circles in Figure~\ref{fig:uvista_numden}. Wild et al. correctly state that all quiescent galaxies must have gone through the post-starburst phase at $z\sim2$ \emph{if} the visibility time for that phase is 250 Myr. Given our high-quality Keck spectra, we can now robustly measure this timescale, which is significantly larger than 250 Myr, and therefore conclude that at most half of the quiescent galaxies at $z\sim2$ were post-starburst at some point in the past. Finally, studies at lower redshift found that the contribution of post-starburst galaxies to the growth of the red sequence declines strongly from $z \sim 1$ to $z \sim 0$ \citep{wild09, dressler13, rowlands18}. This is in qualitative agreement with Figure~\ref{fig:uvista_numden}, which shows that post-starburst galaxies become less important at later cosmic times.


\begin{figure*}[tbp]
\centering
\includegraphics[width=0.7\textwidth]{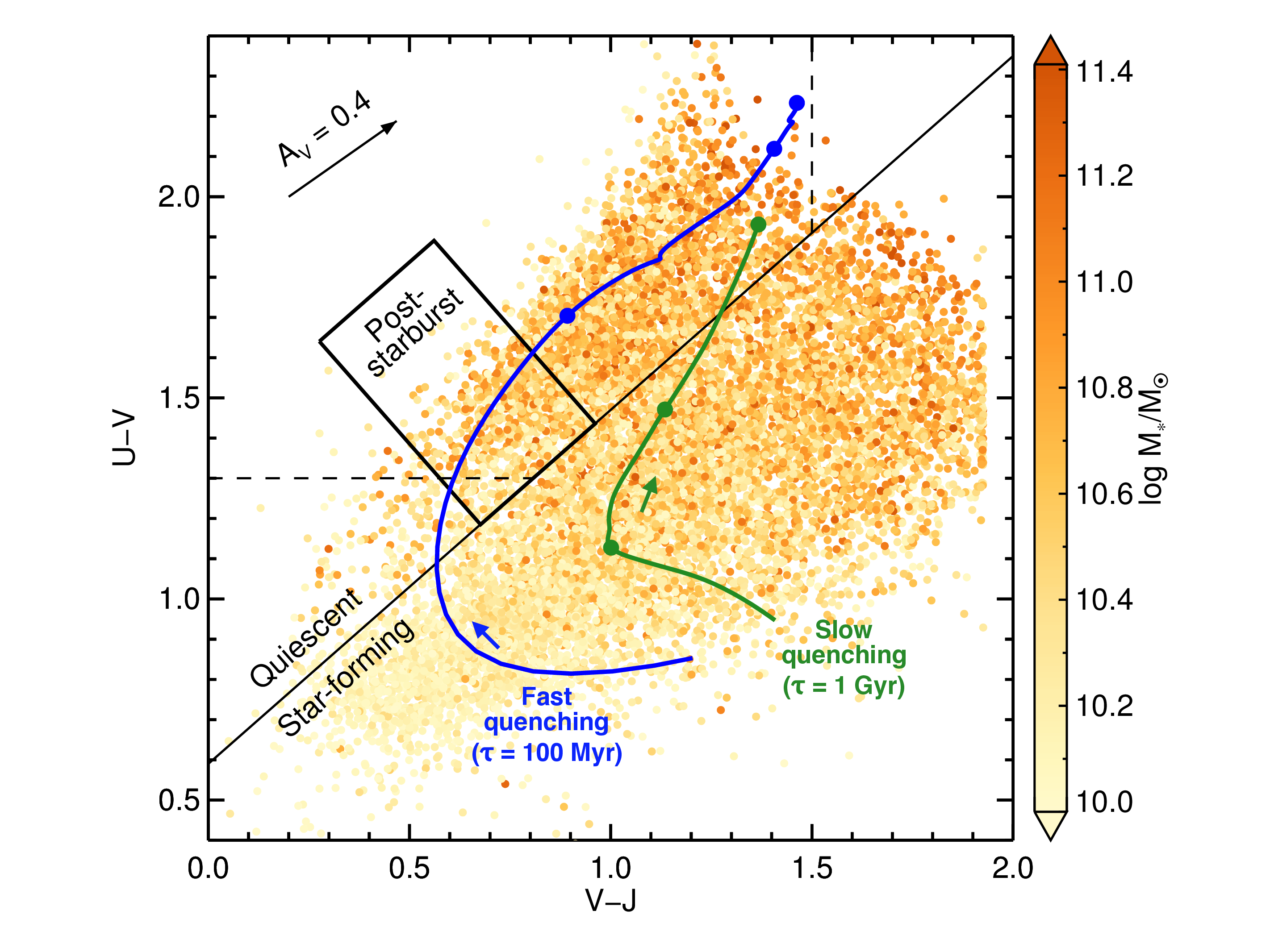}
\caption{Illustration of the fast and slow quenching paths on the $UVJ$ diagram. The distribution of all galaxies in the UltraVISTA catalog with $1.5 < z < 2.0$ and $\log \Mstar/\Msun > 10$ is shown in the background, color-coded by stellar mass. The diagonal line represents our adopted division into star-forming and quiescent galaxies; the dashed lines are additional constraints used by \citet{muzzin13} which we do not use. The black rectangle marks the region that we use to select post-starburst galaxies. The colored lines show two tau models, one with $\tau=100$ Myr (green), the other with $\tau=1$ Gyr (blue); the filled points on the tracks mark even intervals of 1 Gyr. In both cases we adopt a simple model for the dust attenuation, which evolves following the star formation rate, with the constraints of $A_V=2$ at the beginning and $A_V=0.4$ at the end. The arrow in the top left corner shows the effect of dust attenuation on the $UVJ$ colors.}
\label{fig:UVJ_cartoon}
\end{figure*}

\section{Discussion: Fast and Slow Quenching}
\label{sec:quenching}

\subsection{Two Quenching Channels}

Our analysis of the rest-frame colors and number densities demonstrates that not all quiescent galaxies go through the post-starburst phase after being quenched. This implies a variety of quenching timescales: some galaxies undergo a \emph{fast quenching}, becoming post-starburst and then evolving along the red sequence, while the remaining ones must follow a \emph{slow quenching} path that never produces the blue colors typical of post-starburst galaxies. We illustrate these two paths on the $UVJ$ diagram in Figure~\ref{fig:UVJ_cartoon} using two simple toy models. The fast quenching track, shown in blue, is exemplified by a tau model with a short timescale, $\tau = 100$ Myr, while for the slow track, in green, we take $\tau = 1$ Gyr.
We also include the effect of dust reddening, which we model in an identical way for the two tracks: the dust attenuation is $A_V=2$ at the beginning, declines following the evolution of the star formation rate, and then reaches a plateau at $A_V=0.4$, the typical value found for our quiescent galaxies.

The toy models are effective at illustrating the salient features of the fast and slow tracks. Clearly, the red sequence on the $UVJ$ diagram can be joined at any point along its diagonal edge, a region often called the green valley. The exact position of the crossing point will depend on both the quenching timescale and the dust attenuation. The qualitative difference between the two tracks is, however, mainly due to the quenching timescale. For $\tau=1$ Gyr, for example, even a dust-free model will not enter the post-starburst region. In order to reproduce the colors of the most extreme post-starburst galaxies, which are even bluer than our selection box, it is necessary to have little dust and extremely short timescales, $\tau < 100$ Myr. The rarity of these very blue quiescent galaxies, evident in Figure~\ref{fig:UVJ_cartoon}, implies that galaxies move across that region very rapidly, consistent with an extrapolation of our age--color relation.

Our number density analysis leads to a \emph{quantitative} result: in order to explain the observed color distribution of quiescent galaxies, the two quenching channels must be in place simultaneously, and have similar quenching rates at $z\sim2$. This is consistent with the \emph{qualitative} result of our earlier study of the LRIS sample \citep{belli15}, in which we proposed that post-starburst galaxies follow a fast quenching path based on the derived SFHs and on the observation that they have systematically smaller sizes than older quiescent galaxies. Other studies, at both low and high redshift, have reached similar conclusions \citep[e.g.,][]{schawinski14, moutard16, wild16, wu18quenching, forrest18, carnall18}. In these works, the two quenching channels are often characterized not only by different timescales but also by different morphological or structural properties, as we discuss below.

\subsection{Physical Properties of Post-Starburst Galaxies}

Given their importance in the context of galaxy quenching, we want to explore in more detail the properties of post-starburst systems. For this analysis we make use of the catalog from the 3D-HST survey \citep{brammer12}, which covers a much smaller area than UltraVISTA, but is based on high-spatial resolution imaging and grism spectroscopy from \HST. In order to explore a variety of physical properties, we assemble different catalogs that have been publicly released, and that include grism redshifts \citep{momcheva16}, rest-frame colors and stellar population properties \citep{skelton14}, size and morphological properties \citep{vanderwel14}, and measurements of the local overdensities \citep{fossati17}.
Given the limited survey area, the number of post-starburst galaxies is small; to increase the statistics we slightly extend the mass range, selecting all objects with $\log \Msun/\Mstar > 10.6$. Still, we find only 4 galaxies with rest-frame colors in the post-starburst region in the redshift range $1 < z < 1.5$; therefore we limit the analysis to the range $1.5 < z < 2.5$, where the cosmic volume probed is larger. With this selection we obtain a sample of 554 quiescent galaxies, of which 65 fulfill the post-starburst color selection.

\begin{figure}[tbp]
\centering
\includegraphics[width=0.45\textwidth]{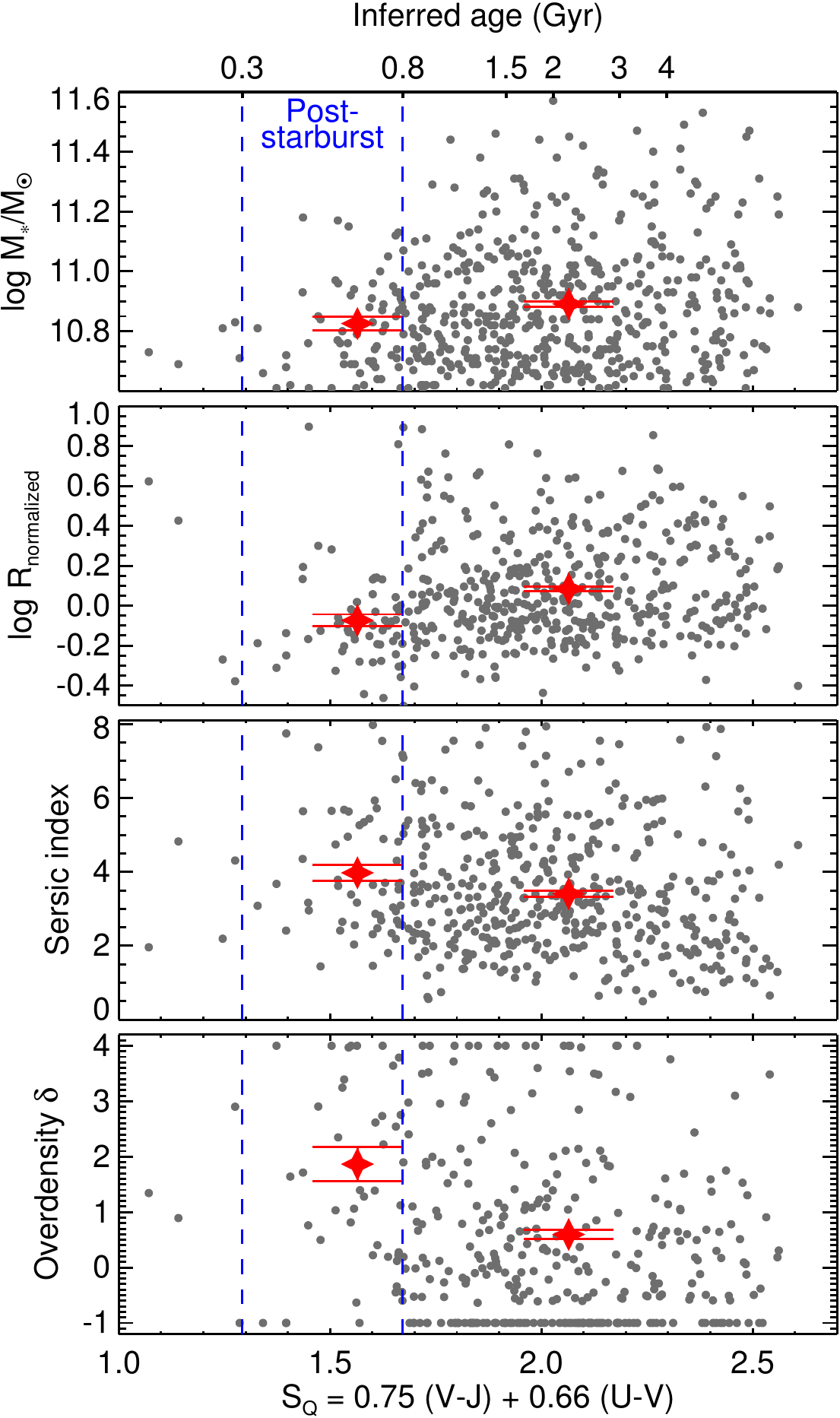}
\caption{Properties of quiescent galaxies from the 3D-HST survey, with $\log \Mstar/\Msun > 10.6$ and $1.5 < z < 2.5$, as a function of their rest-frame color \Sq, which we convert to inferred stellar age on the top axis using Equation~\ref{eq:inferred_age}. From top to bottom, we show the distribution of stellar mass, mass-normalized effective size, Sersic index, and overdensity $\delta$ within a radius of 0.75 Mpc. The average values for the post-starburst population (300 Myr $ < t_{50} < $ 800 Myr) and for the population of old quiescent galaxies ($t_{50} > $ 800 Myr) are shown in red.}
\label{fig:3DHST_properties}
\end{figure}

We plot some of the physical properties for this sample of quiescent galaxies in Figure~\ref{fig:3DHST_properties}, as a function of the rest-frame color \Sq, defined in Equation~\ref{eq:coords}. Using the age--color relation we can then convert this into an inferred stellar age, shown in the top axis. The post-starburst galaxies are those with an inferred age between 300 and 800 Myr.
The figure shows that this population has clearly distinct properties compared to the older quiescent systems. Post-starburst systems tend to have lower stellar masses, consistent with the mass function study performed by \citet{wild16}. To compare the structural properties, we calculate for each object a mass-normalized size by taking the measured effective radius and dividing it by the size typical for quiescent galaxies at that stellar mass and redshift, which we take from the \citet{vanderwel14} mass-size relation.
The distribution of normalized sizes shows that post-starburst galaxies are more compact than the overall quiescent population. This has been first shown by \citet{whitaker12} and successively confirmed by a number of studies \citep{belli15, yano16, almaini17, maltby18, wu18quenching}. The distribution of Sersic indices does not vary strongly with stellar age, with a marginal increase towards the post-starburst region, indicating that these systems are likely spheroidal \citep{almaini17, maltby18}.
Finally, in the last panel we show the distribution of the local overdensity, finding that roughly 80\% of post-starburst galaxies are found in an overdense environment ($\delta > 0$), as opposed to 25\% of the older quiescent galaxies. Despite residing in overdensities, post-starburst galaxies are not likely to be satellites: using the \citet{fossati17} catalog we find that three quarters of the post-starburst galaxies shown in Figure~\ref{fig:3DHST_properties} are centrals with a probability larger than 90\%. Our results are consistent with the excess of post-starburst galaxies that has been found in clusters at $z<1$ \citep{socolovsky18, moutard18}. However, these post-starburst galaxies at lower redshifts are less massive ($\log \Mstar/\Msun < 10.5$), and are likely to be satellites.

The emerging picture is one in which the fast quenching channel produces post-starburst galaxies that are preferentially compact, spheroidal, have lower stellar masses and lie in higher density environment compared to the bulk of the quiescent population. The fact that post-starburst galaxies have such distinct physical properties points to a physical difference between fast and slow quenching, rather than to a single physical process characterized by a wide distribution of quenching timescales. One pressing question remains: What are the physical mechanisms responsible for the two quenching channels?

\subsection{Fast Quenching}

Prime candidates for the fast quenching mechanism are gas-rich processes, that naturally explain both the short timescales and the compact sizes found in post-starburst galaxies. Major mergers \citep{hopkins06mergers} or violent disk instabilities \citep{dekel14} can drive nuclear inflows and cause centrally concentrated starbursts. These processes have been explored in detail in recent theoretical work, specifically because they can explain why massive quiescent galaxies at high redshift are generally very small compared to their local counterparts. Using a suite of hydrodynamical simulations, \citet{zolotov15} showed that mergers or violent disk instabilities can trigger a \emph{compaction} event in which a massive stellar core is formed in a short time, with a resulting decrease in the galaxy effective size. Quenching follows naturally, due to a combination of gas exhaustion and stellar feedback; AGN feedback may also be involved, particularly for the maintenance of quiescence over longer timescales. However, these simulations predict a combined timescale for compaction and quenching of the order of 2 Gyr \citep[see also][]{tacchella16compaction}, which is substantially longer than what we measure for the post-starburst objects, and more similar to our slow quenching channel. This discrepancy may indicate the need for stronger feedback or additional physical mechanisms. A hint for the important role played by AGN feedback in this critical phase comes from the high incidence of ionized gas emission (not due to star formation) found in the post-starburst region of the $UVJ$ diagram \citep{belli17kmos3d}.

Compaction can be triggered by a number of different processes, including major or minor mergers, counter-rotating streams, perturbation due to giant clumps, and tidal interactions \citep{zolotov15}. The large fraction of post-starburst galaxies residing in overdensities suggests a prominent role for mergers and interactions with nearby galaxies \citep[see also, e.g.,][]{brodwin13,wu14}. We note that post-starburst systems are mostly centrals and therefore are not affected by the so-called environmental quenching, which has been proposed for explaining the abundance of quiescent low-mass systems in dense environments, and applies only to satellites \citep{peng10}. A link with environmental quenching, however, cannot be ruled out at $z<1$, where the post-starburst population consists of lower-mass galaxies \citep{maltby18}.

Compaction is not the only possible explanation for the small sizes of quiescent galaxies at high redshift. The alternative possibility is that these systems formed at very high redshift, when star-forming disks were significantly smaller, and then evolved passively retaining their compact structures \citep{wellons15}. However, the young ages we measure for post-starburst galaxies rule out this scenario. We conclude that the observed properties of post-starburst galaxies require a compaction phenomenon, defined in its broadest sense as the formation of quiescent galaxies that are more compact than their star-forming progenitors. To establish exactly how and why this happens, more theoretical and observational studies are clearly needed.

In Section~\ref{sec:numdens} we found that the importance of the fast quenching channel increases with redshift, up to $z\sim2.5$. At even higher redshift it is likely that fast quenching represents, in practice, the \emph{only} available track, simply because of the young age of the universe.
The earliest quiescent galaxy with a spectroscopic confirmation is a massive system at $z=3.717$, that features strong Balmer absorption lines, typical of post-starburst systems \citep{glazebrook17}. Interestingly, this object happens to have a massive close companion \citep{simpson17, schreiber18jekyll}. It is possible that the very first massive quiescent galaxies, which reside in overdense environments, were all quenched rapidly via major mergers.

\subsection{Slow Quenching}

Normal star forming galaxies have depletion times of about 1 Gyr \citep[e.g.,][]{genzel10}, which means that a continuous replenishment of the gas reservoir, via gas inflows, is needed to continue the star formation activity over a cosmological timescale. If such replenishment is halted for any reason, the galaxy slowly consumes the remaining cold gas and quenches over long timescales. Such gas starvation is therefore the simplest explanation for the slow quenching path. Unfortunately, there are many different physical mechanisms that can produce gas starvation.
The most commonly invoked scenario is one where the gas cannot cool when the halo grows beyond a threshold mass due to the formation of a stable virial shock \citep{birnboim03, keres05}, possibly in conjunction with heating by AGN radio-mode feedback \citep{croton06, bower06}. There are, however, other possibilities: the gas may cool, but then made stable against collapse by the deep potential well of a large stellar bulge \citep{martig09, genzel14quenching}; or the gas inflows could stop even before reaching the halo \citep{feldmann15}.

In the local universe, the stellar metallicity of quiescent galaxies is substantially larger than that of star-forming objects of the same mass, which is a signature of starvation \citep{peng15}. This is consistent with the increasing importance of the slow quenching channel at later cosmic epochs.

\begin{figure*}[htbp]
  \vspace{3mm}
\centering
\includegraphics[width=1.0\textwidth]{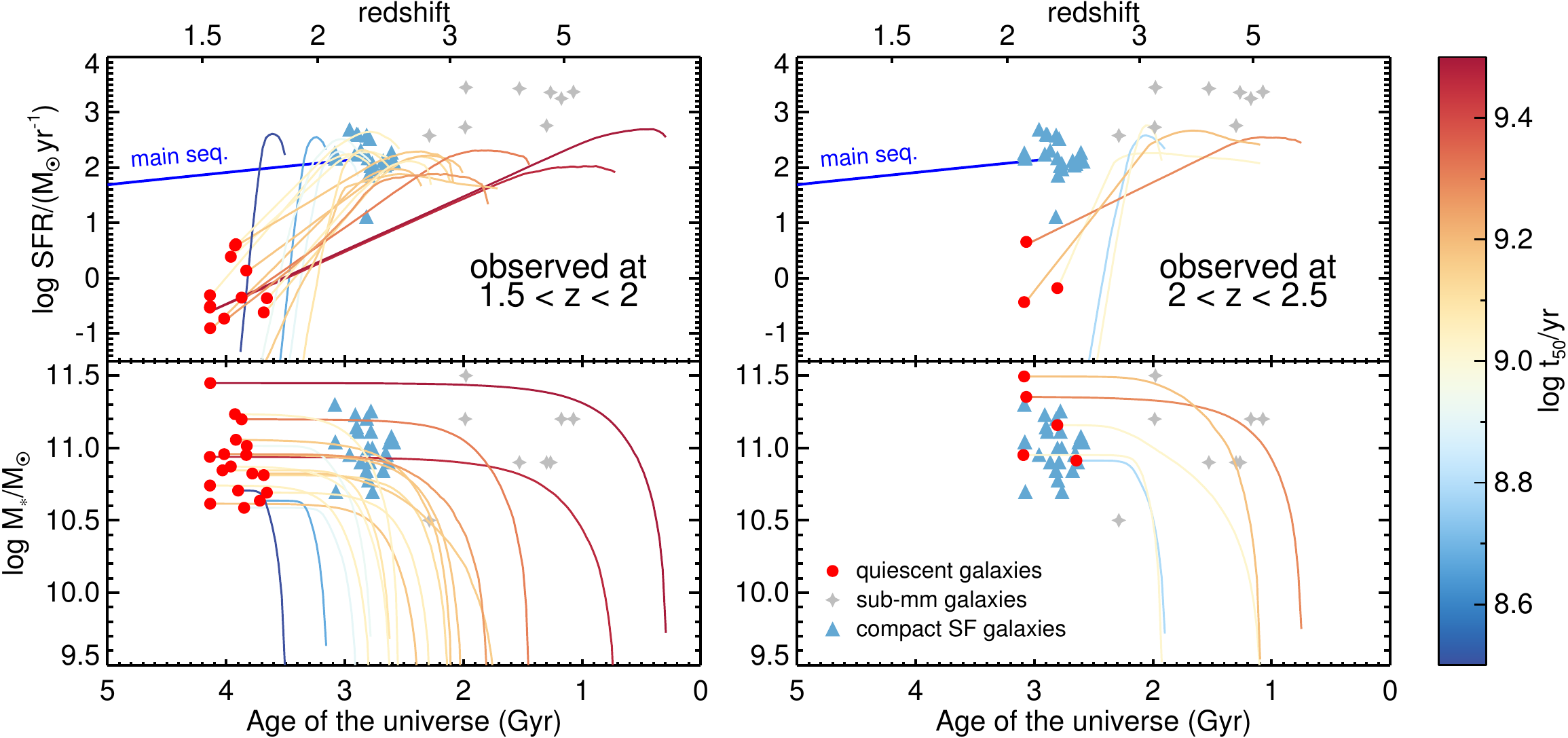}
\caption{Reconstructed SFHs (top) and stellar mass growth (bottom) for the quiescent galaxies in our MOSFIRE sample, adopting the constant + tau model. The left and right panels show two redshift bins, corresponding to galaxies observed at $1.5 < z < 2$ and $2 < z < 2.5$ respectively. The lines are colored according to the median stellar age $t_{50}$, and the red points mark the observed redshift. In the top panels, the blue line indicates the evolution of the main sequence for a fixed stellar mass of $10^{11}\Msun$, from \citet{whitaker14}. The sample of sub-mm galaxies from \citet{toft14} is shown as gray diamonds, while the compact star-forming galaxies from \citet{vandokkum15} are shown as blue triangles.}
\label{fig:mass_assembly}
\end{figure*}

\subsection{Identifying the Star-Forming Progenitors}

Another way to constrain the quenching mechanisms is to link quiescent galaxies with their star-forming progenitors. Using the star formation histories from our spectral fits, we can estimate physical properties such as stellar mass and star formation rate for the progenitors of the quiescent galaxies in our sample. We caution that this analysis is strongly affected by the assumption of the SFH model, and should be considered as purely qualitative.

In Figure \ref{fig:mass_assembly} we show the inferred past evolution in both star formation rate (top) and stellar mass (bottom) for our MOSFIRE sample, split into two redshift bins. The tracks shown in the figure correspond to the constant + tau model for the SFH, and are color-coded according to the median stellar age. We compare the inferred evolution with the properties of spectroscopically confirmed sub-mm galaxies \citep{toft14} and compact star-forming galaxies \citep{vandokkum15}, two populations that have been proposed as possible progenitors of massive quiescent galaxies. The masses and star formation rates of compact star-forming galaxies are consistent with those of the progenitors of quiescent galaxies, at least for those observed at $1.5 < z < 2$, and are also consistent with the star-forming main sequence \citep[e.g.,][]{whitaker14}. On the other hand, the sub-mm galaxies have substantially higher values of star formation rates, which are not consistent with the SFH tracks. However this discrepancy can be easily explained by the typical duty cycle of sub-mm galaxies, which may be as short as 40 Myr \citep{toft14}: such a short-lived phase would leave virtually no signature on the spectra observed more than a Gyr later.
Figure~\ref{fig:mass_assembly} also shows that the peak value of star formation rate is roughly the same for all quiescent galaxies, around 100 -- 300 \Msun/yr, and does not depend on the median stellar age. This means that the SFHs of post-starburst galaxies do not require a \emph{burst} of star formation above the main sequence; they simply require rapid formation and quenching.

In the compaction scenario, galaxies first shrink and then quench. The population of compact star-forming galaxies is thought to represent the intermediate stage after compaction and before quenching \citep{barro13, vandokkum15}. Measurements of the kinematics have been used to draw a link between compact star-forming galaxies and quiescent systems. The integrated velocity dispersion of the gas in compact star-forming galaxies is in broad agreement with that of the stellar populations in quiescent galaxies \citep{barro14mosfire, nelson14}. A comparison of the spatially resolved kinematics, though, would be even more compelling. Recent observations showed that the gas in compact star-forming galaxies is actually rotating rapidly, both in the ionized \citep{vandokkum15, wisnioski18}, and molecular components \citep{tadaki17CO, barro17CO}. This is not in agreement with the kinematics of the most massive quiescent galaxies at $z\sim0$, which tend to be pressure-supported. At high redshift, however, quiescent galaxies seem to be very different from local ellipticals.

Our previous analysis of the MOSFIRE sample, which focused on the stellar kinematics, suggested that the rotational support of quiescent galaxies is significantly larger at high redshift \citepalias{belli17mosfire}. This surprising result is consistent with detailed studies of a few gravitationally lensed systems at $z>2$ \citep{newman15, newman18II, toft17}, and a study of a large sample at $z<1$ \citep{bezanson18}.
The fact that massive quiescent galaxies at high redshift present a large degree of rotational support has two important consequences in the context of quenching. First, it suggests that the giant ellipticals observed at $z\sim0$ were originally quiescent disks, which lost their angular momentum likely because of dry mergers. This means that, contrary to the standard view, the morphological transformation does not need to happen simultaneously with the quenching process in order to form ellipticals. Secondly, it offers an additional constraint on the quenching process, which needs to produce a rotating stellar system. Since starvation leaves the morphology and kinematics untouched, the slow quenching channel should produce preferentially rotating quiescent systems. However, gas-rich mergers can also produce rotating remnants \citep{wuyts10}. It is therefore not clear whether kinematics can be used for a clean identification of the progenitors.

The question of whether the two quenching channels require distinct populations of progenitors is still open. It is possible that all star-forming galaxies can potentially undergo fast quenching after a close interaction or merger. On the other hand, post-starburst galaxies show smaller stellar masses compared to older quiescent systems, and the tracks shown in Figure~\ref{fig:UVJ_cartoon} also indicate a mass segregation between fast quenching and slow quenching. If this were the case, star-forming galaxies above a certain stellar mass (which depends on redshift) would be the ideal candidates to be progenitors of the slow quenching track.


\section{Summary and Conclusions}
\label{sec:summary}

\subsection{The SFHs of $z\sim2$ Massive Quiescent Galaxies}

We analyzed the deep near-infrared spectra of 24 massive quiescent galaxies at $1.5< z < 2.5$ obtained with Keck/MOSFIRE. The high quality of the spectroscopic data, which we now release publicly, allowed us to perform a study of the stellar population properties of these objects to an unprecedented level of detail.

We took full advantage of all the available data by fitting templates simultaneously to the observed spectroscopy and broadband photometry. We explored six different SFH models with the goal of understanding how this impacts the derived stellar population properties. We found that the SSP and the top-hat models are not able to reproduce the observed data for at least some galaxies in our sample. The remaining four models were all equally successful, including the commonly used tau model which has only two free parameters. Despite the degeneracy among different SFH models, the median stellar ages are found to be robust, together with the sSFR averaged over the past 100 Myr. However, the duration of the star formation depends strongly on the SFH shape, and therefore cannot be currently measured robustly.

In order to break the degeneracy between different SFH models, significantly deeper spectra would be needed. This may be possible for individual objects, but for larger samples we will have to wait the next generation of ground-based telescopes. A complementary approach is to use the stellar metallicity to indirectly probe the formation timescale \citep{kriek16}, which however requires a wider wavelength range than that available for our sample.

\subsection{Ionized Gas in Quiescent Galaxies}

By stacking shallower spectra obtained at redder wavelengths, we detected the \Halpha\ and \NII\ emission lines. The \Halpha\ line is faint, and would not be detected without a proper subtraction of the underlying stellar absorption. We measured a high line flux ratio, \NII/H$\alpha = 1.0 \pm 0.1$, which rules out star formation as the sole origin of the ionization. About 10\% of quiescent galaxies at high redshift feature emission lines with a similar flux ratio, which is likely due to AGN-driven outflows \citep{belli17kmos3d}. Our finding suggests that low-level AGN activity is even more common in quiescent galaxies, but in most cases is too weak to be noticed. This is also confirmed by the sample of gravitationally lensed quiescent galaxies in \citet{newman18I}, where the lensing magnification allows to observe faint emission lines in most of the objects. Such low-level AGN feedback may play a role in either quenching the star formation or in maintaining the galaxies quiescent.

Assuming that the dust-corrected \Halpha\ flux is entirely due to star formation yields a strict upper limit on the star formation rate, which is $0.9 \pm 0.1$ \Msun/yr. This corresponds to an average $\log \mathrm{sSFR} \sim -11.1$, confirming the results of the spectral fitting. This low value of star formation is in broad agreement with other studies that used less direct methods \citep{fumagalli14, man16, gobat17}, and suggests a very effective quenching in massive galaxies already at $z\sim2$.

\subsection{The Age--Color Relation}

By combining the MOSFIRE observations with our previous LRIS sample at lower redshift, we found that the majority of quiescent galaxies at $1 < z < 2.5$ lie on a remarkably tight relation between stellar age and $UVJ$ colors, which is independent of redshift and stellar mass. Such a relation deviates from that predicted by a simple model of passive evolution due to the contribution of dust reddening, which evolves rapidly during the post-starburst phase. The use of a large sample of high-quality spectra is therefore critical for determining the age--color relation. We quantified this relation and its scatter, which is only 0.13 dex in age, and we gave an explicit calibration that can be used to infer the approximate stellar age of any high-redshift quiescent galaxy given their rest-frame colors, without the need for spectroscopic data. This is a simple yet powerful method that allowed us to model the evolution of the quiescent population, and that can also be used for other applications, such as selecting specific targets for spectroscopic follow-up.

Based on our observations and on previous studies \citep[e.g.,][]{whitaker15, merlin18}, we argued that the lower and upper edges commonly used to define the quiescent region in the $UVJ$ diagram are not justified, and artificially exclude very young and very dusty quiescent galaxies. The young quiescent galaxies, also called post-starburst, represent a key evolutionary phase in the formation of the red sequence. The nature of the dusty objects (which are the only galaxies in our sample that do not lie on the age--color relation), instead, is currently not known: it is possible that they consists of edge-on disks and/or interacting systems.

It would be interesting to extend the age--color relation to higher redshifts, however the few spectroscopic measurements available at $z>3$ show that $UVJ$ colors may become less effective in dividing quiescent and star-forming galaxies \citep{schreiber18}.

\subsection{Galaxy Quenching}

Combining the age--color relation, which we derived from our spectroscopic sample, with a large, volume-limited sample drawn from the UltraVISTA survey, we were able to model the evolution of the red sequence at $1<z<2.5$ in a quantitative and self-consistent way. The main result of this analysis is the identification of two quenching channels: fast quenching produces compact post-starburst objects, while slow quenching contributes to the growth of the red sequence by adding objects with stellar populations that are already old. Fast quenching is increasingly more important at high redshift, contributing for about half of the growth of the quiescent population at $z\sim2$.

Post-starburst objects are clearly distinct from the bulk of the red sequence, and feature lower stellar masses, compact sizes, spheroidal morphologies, and higher overdensities, as also shown by a number of recent studies. Together with their rapid formation timescale, these porperties fit well a scenario in which violent, gas-rich events such as mergers are responsible for their formation. The slow quenching channel, instead, must be due to a smooth decline in the gas reservoir, which may be caused by fuel starvation. However, there are many physical processes that may be responsible for starvation, including AGN feedback, virial shock heating, cosmological or gravitational effects. \\

It is a pleasure to acknowledge Omar Almaini, Nir Mandelker, Stefano Andreon, Philipp Plewa, and Iary Davidzon for insightful discussions. We also thank the anonymous referee for constructive comments.
The authors recognize and acknowledge the very significant cultural role and reverence that the summit of Mauna Kea has always had within the indigenous Hawaiian community. We are most fortunate to have the opportunity to conduct observations from this mountain.


\bibliography{sirio}



\appendix
\section{Results of the Spectral Fitting for the LRIS Sample}
\label{sec:appendix_lris}

\begin{deluxetable*}{llcccccc}
\tabletypesize{\footnotesize}
\tablewidth{0pc}
\tablecaption{Stellar Population Properties of the LRIS Sample \label{tab:sample_lris}}
\tablehead{
\colhead{ID} & \colhead{Field} & \colhead{z} & \colhead{$t_{50}$\tablenotemark{a}} &  \colhead{$A_V$\tablenotemark{a}} &  \colhead{$Z$\tablenotemark{a}} &  \colhead{log \Mstar/\Msun\tablenotemark{a}} &  \colhead{log sSFR\tablenotemark{a,b}}
\\
(3D-HST) & & & $10^9$ yr & mag & & & yr$^{-1}$
}
\startdata
17678 & GOODS-N & 1.60 & 0.77 $ \pm $ 0.07 & 0.52 $ \pm $ 0.07 & 0.013 $ \pm $ 0.002 & 10.84 $ \pm $ 0.01 & -11.8 $ \pm $ 3.7 \\
23303 & COSMOS & 1.33 & 0.91 $ \pm $ 0.05 & 0.55 $ \pm $ 0.05 & 0.013 $ \pm $ 0.002 & 10.69 $ \pm $ 0.01 & -14.7 $ \pm $ 6.9 \\
33158 & EGS & 1.40 & 1.0 $ \pm $ 0.2 & 0.76 $ \pm $ 0.12 & 0.016 $ \pm $ 0.003 & 10.92 $ \pm $ 0.02 & -9.9 $ \pm $ 0.1 \\
28523 & COSMOS & 1.58 & 1.0 $ \pm $ 0.05 & 0.62 $ \pm $ 0.04 & 0.015 $ \pm $ 0.002 & 11.33 $ \pm $ 0.01 & -24.9 $ \pm $ 8.6 \\
4748 & GOODS-N & 1.27 & 1.0 $ \pm $ 0.1 & 0.44 $ \pm $ 0.05 & 0.024 $ \pm $ 0.003 & 10.78 $ \pm $ 0.03 & -22.3 $ \pm $ 9.0 \\
17533 & EGS & 1.26 & 1.0 $ \pm $ 0.08 & 0.59 $ \pm $ 0.05 & 0.016 $ \pm $ 0.003 & 10.69 $ \pm $ 0.02 & -10.4 $ \pm $ 0.06 \\
21628 & COSMOS & 1.43 & 1.0 $ \pm $ 0.1 & 0.69 $ \pm $ 0.09 & 0.016 $ \pm $ 0.003 & 10.60 $ \pm $ 0.02 & -11.0 $ \pm $ 2.4 \\
1433 & GOODS-N & 1.40 & 1.1 $ \pm $ 0.1 & 0.31 $ \pm $ 0.09 & 0.016 $ \pm $ 0.005 & 10.52 $ \pm $ 0.02 & -10.2 $ \pm $ 0.07 \\
4275 & COSMOS & 1.22 & 1.1 $ \pm $ 0.09 & 0.61 $ \pm $ 0.07 & 0.014 $ \pm $ 0.004 & 10.79 $ \pm $ 0.02 & -10.5 $ \pm $ 0.05 \\
41108 & GOODS-S & 1.22 & 1.2 $ \pm $ 0.4 & 0.12 $ \pm $ 0.05 & 0.020 $ \pm $ 0.004 & 10.13 $ \pm $ 0.05 & -11.3 $ \pm $ 2.5 \\
22782 & EGS & 1.08 & 1.3 $ \pm $ 0.2 & 0.22 $ \pm $ 0.06 & 0.018 $ \pm $ 0.004 & 10.68 $ \pm $ 0.04 & -11.4 $ \pm $ 4.8 \\
36048 & EGS & 1.28 & 1.3 $ \pm $ 0.1 & 0.48 $ \pm $ 0.05 & 0.018 $ \pm $ 0.004 & 10.90 $ \pm $ 0.03 & -10.4 $ \pm $ 0.05 \\
12167 & EGS & 1.19 & 1.4 $ \pm $ 0.3 & 0.36 $ \pm $ 0.06 & 0.016 $ \pm $ 0.004 & 10.79 $ \pm $ 0.04 & -13.1 $ \pm $ 3.8 \\
19419 & COSMOS & 1.26 & 1.4 $ \pm $ 0.3 & 0.27 $ \pm $ 0.06 & 0.011 $ \pm $ 0.002 & 10.49 $ \pm $ 0.04 & -16.6 $ \pm $ 8.2 \\
23909 & EGS & 1.22 & 1.5 $ \pm $ 0.2 & 0.22 $ \pm $ 0.06 & 0.015 $ \pm $ 0.005 & 10.54 $ \pm $ 0.04 & -11.3 $ \pm $ 1.0 \\
9636 & COSMOS & 1.08 & 1.6 $ \pm $ 0.2 & 0.42 $ \pm $ 0.07 & 0.018 $ \pm $ 0.004 & 10.26 $ \pm $ 0.02 & -10.2 $ \pm $ 0.06 \\
796 & COSMOS & 1.15 & 1.6 $ \pm $ 0.2 & 1.12 $ \pm $ 0.06 & 0.020 $ \pm $ 0.003 & 11.12 $ \pm $ 0.03 & -10.6 $ \pm $ 0.07 \\
16571 & COSMOS & 1.10 & 1.6 $ \pm $ 0.2 & 0.95 $ \pm $ 0.09 & 0.026 $ \pm $ 0.005 & 10.65 $ \pm $ 0.04 & -10.7 $ \pm $ 0.08 \\
30221 & COSMOS & 1.24 & 1.7 $ \pm $ 0.3 & 0.96 $ \pm $ 0.06 & 0.011 $ \pm $ 0.003 & 10.84 $ \pm $ 0.03 & -28.2 $ \pm $ 16.8 \\
8962 & COSMOS & 1.09 & 1.8 $ \pm $ 0.4 & 0.97 $ \pm $ 0.06 & 0.013 $ \pm $ 0.003 & 10.53 $ \pm $ 0.04 & -12.4 $ \pm $ 2.5 \\
23084 & COSMOS & 1.04 & 1.8 $ \pm $ 0.2 & 0.22 $ \pm $ 0.04 & 0.011 $ \pm $ 0.002 & 10.54 $ \pm $ 0.02 & -11.3 $ \pm $ 0.08 \\
30145 & COSMOS & 1.40 & 1.8 $ \pm $ 0.3 & 0.34 $ \pm $ 0.07 & 0.013 $ \pm $ 0.003 & 10.76 $ \pm $ 0.03 & -15.4 $ \pm $ 12.7 \\
19719 & COSMOS & 1.14 & 1.8 $ \pm $ 0.2 & 0.53 $ \pm $ 0.05 & 0.014 $ \pm $ 0.003 & 10.94 $ \pm $ 0.02 & -10.2 $ \pm $ 0.05 \\
19796 & COSMOS & 1.26 & 1.8 $ \pm $ 0.09 & 0.28 $ \pm $ 0.03 & 0.009 $ \pm $ 0.001 & 10.99 $ \pm $ 0.02 & -11.5 $ \pm $ 0.06 \\
28619 & COSMOS & 1.25 & 1.9 $ \pm $ 0.2 & 0.62 $ \pm $ 0.04 & 0.010 $ \pm $ 0.003 & 10.85 $ \pm $ 0.02 & -11.3 $ \pm $ 0.08 \\
25627 & COSMOS & 1.10 & 1.9 $ \pm $ 0.3 & 0.15 $ \pm $ 0.04 & 0.018 $ \pm $ 0.002 & 11.01 $ \pm $ 0.04 & -15.0 $ \pm $ 4.3 \\
20342 & COSMOS & 1.10 & 2.0 $ \pm $ 0.2 & 0.61 $ \pm $ 0.07 & 0.014 $ \pm $ 0.003 & 11.05 $ \pm $ 0.02 & -10.5 $ \pm $ 0.07 \\
38867 & EGS & 1.11 & 2.1 $ \pm $ 0.2 & 0.41 $ \pm $ 0.04 & 0.018 $ \pm $ 0.003 & 10.88 $ \pm $ 0.02 & -10.7 $ \pm $ 0.04 \\
26884 & EGS & 1.24 & 2.3 $ \pm $ 0.1 & 0.58 $ \pm $ 0.04 & 0.012 $ \pm $ 0.002 & 11.15 $ \pm $ 0.02 & -11.2 $ \pm $ 0.06 \\
27877 & EGS & 1.05 & 2.3 $ \pm $ 0.3 & 0.53 $ \pm $ 0.05 & 0.014 $ \pm $ 0.003 & 11.00 $ \pm $ 0.02 & -11.6 $ \pm $ 0.7 \\
5087 & EGS & 1.41 & 2.4 $ \pm $ 0.3 & 0.37 $ \pm $ 0.06 & 0.010 $ \pm $ 0.002 & 10.91 $ \pm $ 0.03 & -11.2 $ \pm $ 0.07 \\
10703 & COSMOS & 1.26 & 2.4 $ \pm $ 0.3 & 0.47 $ \pm $ 0.04 & 0.012 $ \pm $ 0.003 & 11.11 $ \pm $ 0.02 & -11.4 $ \pm $ 0.07 \\
17070 & GOODS-N & 1.24 & 2.4 $ \pm $ 0.1 & 0.28 $ \pm $ 0.03 & 0.009 $ \pm $ 0.001 & 10.96 $ \pm $ 0.01 & -30.6 $ \pm $ 22.8 \\
42466 & GOODS-S & 1.42 & 2.5 $ \pm $ 0.3 & 0.19 $ \pm $ 0.06 & 0.012 $ \pm $ 0.002 & 10.92 $ \pm $ 0.02 & -12.4 $ \pm $ 8.5 \\
8381 & COSMOS & 1.41 & 2.5 $ \pm $ 0.4 & 0.45 $ \pm $ 0.07 & 0.010 $ \pm $ 0.002 & 10.92 $ \pm $ 0.03 & -10.3 $ \pm $ 0.08 \\
30838 & COSMOS & 1.32 & 2.5 $ \pm $ 0.4 & 0.51 $ \pm $ 0.06 & 0.011 $ \pm $ 0.002 & 10.69 $ \pm $ 0.03 & -11.2 $ \pm $ 0.07 \\
20106 & EGS & 1.06 & 2.6 $ \pm $ 0.5 & 0.93 $ \pm $ 0.08 & 0.013 $ \pm $ 0.004 & 11.07 $ \pm $ 0.04 & -10.3 $ \pm $ 0.09 \\
19083 & COSMOS & 1.26 & 2.6 $ \pm $ 0.2 & 0.22 $ \pm $ 0.04 & 0.009 $ \pm $ 0.001 & 10.93 $ \pm $ 0.01 & -11.2 $ \pm $ 0.04 \\
43042 & GOODS-S & 1.42 & 2.6 $ \pm $ 0.5 & 0.77 $ \pm $ 0.05 & 0.009 $ \pm $ 0.005 & 11.33 $ \pm $ 0.03 & -33.6 $ \pm $ 26.1 \\
4527 & COSMOS & 1.26 & 2.7 $ \pm $ 0.3 & 0.43 $ \pm $ 0.06 & 0.014 $ \pm $ 0.003 & 11.01 $ \pm $ 0.02 & -12.3 $ \pm $ 10.1 \\
2543 & COSMOS & 1.33 & 2.8 $ \pm $ 0.3 & 0.21 $ \pm $ 0.07 & 0.012 $ \pm $ 0.002 & 11.09 $ \pm $ 0.02 & -11.9 $ \pm $ 2.5 \\
16164 & EGS & 1.19 & 2.9 $ \pm $ 0.4 & 0.58 $ \pm $ 0.08 & 0.018 $ \pm $ 0.004 & 10.95 $ \pm $ 0.04 & -10.5 $ \pm $ 0.07 \\
11783 & COSMOS & 1.26 & 2.9 $ \pm $ 0.3 & 0.27 $ \pm $ 0.03 & 0.017 $ \pm $ 0.003 & 11.13 $ \pm $ 0.02 & -11.2 $ \pm $ 0.04 \\
27472 & EGS & 1.22 & 3.1 $ \pm $ 0.5 & 0.15 $ \pm $ 0.06 & 0.015 $ \pm $ 0.004 & 10.68 $ \pm $ 0.04 & -11.6 $ \pm $ 0.1 \\
13041 & GOODS-N & 1.32 & 3.2 $ \pm $ 0.4 & 0.41 $ \pm $ 0.05 & 0.012 $ \pm $ 0.002 & 11.09 $ \pm $ 0.03 & -11.6 $ \pm $ 0.09 \\
14060 & COSMOS & 1.19 & 3.2 $ \pm $ 0.2 & 0.06 $ \pm $ 0.03 & 0.019 $ \pm $ 0.002 & 10.85 $ \pm $ 0.02 & -12.1 $ \pm $ 0.1 \\
30393 & EGS & 1.18 & 3.4 $ \pm $ 0.5 & 0.17 $ \pm $ 0.04 & 0.010 $ \pm $ 0.002 & 11.09 $ \pm $ 0.05 & -11.8 $ \pm $ 0.09 \\
31555 & COSMOS & 1.32 & 3.5 $ \pm $ 0.3 & 0.27 $ \pm $ 0.05 & 0.013 $ \pm $ 0.002 & 11.22 $ \pm $ 0.03 & -11.1 $ \pm $ 0.05 \\
31613 & GOODS-S & 1.22 & 3.5 $ \pm $ 0.5 & 0.09 $ \pm $ 0.05 & 0.012 $ \pm $ 0.002 & 10.81 $ \pm $ 0.04 & -11.8 $ \pm $ 0.08 \\
22126 & EGS & 1.06 & 3.6 $ \pm $ 0.6 & 0.37 $ \pm $ 0.09 & 0.014 $ \pm $ 0.003 & 11.37 $ \pm $ 0.05 & -11.9 $ \pm $ 2.7 \\
16822 & EGS & 1.26 & 3.6 $ \pm $ 0.4 & 0.25 $ \pm $ 0.05 & 0.010 $ \pm $ 0.002 & 10.99 $ \pm $ 0.04 & -11.6 $ \pm $ 0.07 \\
18045 & EGS & 1.01 & 3.8 $ \pm $ 0.6 & 0.84 $ \pm $ 0.08 & 0.012 $ \pm $ 0.003 & 11.33 $ \pm $ 0.05 & -11.1 $ \pm $ 0.08 \\
24456 & EGS & 1.10 & 3.9 $ \pm $ 0.4 & 0.09 $ \pm $ 0.04 & 0.017 $ \pm $ 0.002 & 11.16 $ \pm $ 0.04 & -12.0 $ \pm $ 0.08 \\
32114 & EGS & 1.03 & 4.0 $ \pm $ 0.5 & 0.12 $ \pm $ 0.05 & 0.016 $ \pm $ 0.002 & 11.04 $ \pm $ 0.04 & -11.9 $ \pm $ 0.09 \\
29863 & EGS & 1.12 & 4.1 $ \pm $ 0.3 & 0.27 $ \pm $ 0.05 & 0.024 $ \pm $ 0.003 & 11.11 $ \pm $ 0.03 & -11.4 $ \pm $ 0.05 \\
30244 & COSMOS & 1.01 & 4.3 $ \pm $ 0.3 & 0.16 $ \pm $ 0.04 & 0.010 $ \pm $ 0.001 & 11.08 $ \pm $ 0.02 & -10.6 $ \pm $ 0.03
\enddata
\tablecomments{More properties, including coordinates, morphological and dynamical measurements, are available in Table 2 of \citet{belli14lris}.}
\tablenotetext{a}{Derived from spectral fitting adopting a constant + tau model. The value and uncertainty are given by, respectively, the median and standard deviation of the posterior distribution.}
\tablenotetext{b}{Averaged over the past 100 Myr.}
\vspace{-5mm}
\end{deluxetable*}


\begin{figure}[htbp]
\centering
\vspace{5mm}
\includegraphics[width=0.95\textwidth]{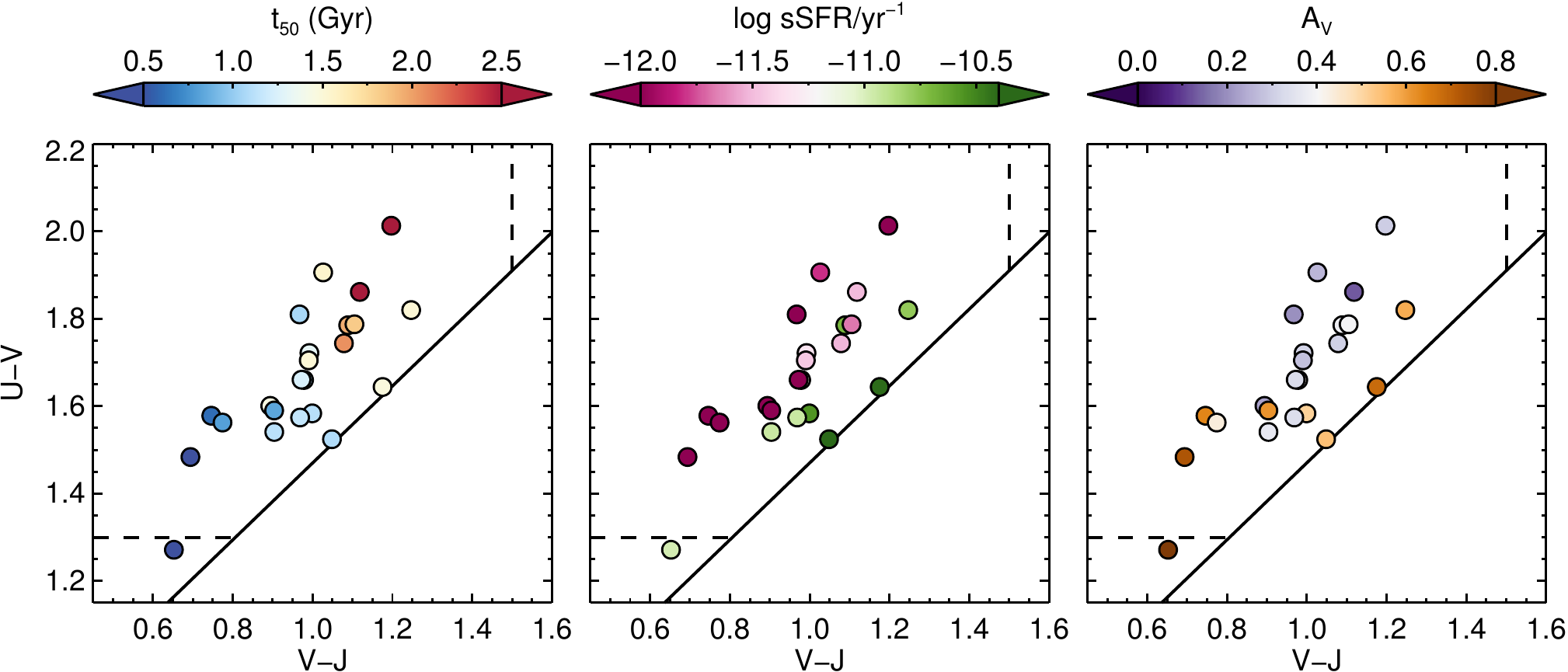}
\caption{Distribution on the $UVJ$ diagram of the median stellar age $t_{50}$, the sSFR averaged over the last 100 Myr, and the dust attenuation $A_V$ for the MOSFIRE sample. The stellar population properties are derived from the spectral fits assuming the constant + tau model; the rest-frame colors are from the 3D-HST catalog. The black line marks the separation between quiescent and star-forming galaxies, and the dashed lines are the additional selection criteria used by \citet{muzzin13}.}
\label{fig:mosfire_UVJ}
\end{figure}

\clearpage

\section{Stellar Populations of the MOSFIRE Sample on the $UVJ$ Diagram}
\label{sec:appendix_uvj}

In Figure~\ref{fig:mosfire_UVJ} we explore the distribution of the stellar population properties of the MOSFIRE sample on the $UVJ$ diagram. The stellar population properties are derived assuming a constant + tau model, while the rest-frame colors are taken from the 3D-HST catalog, which are therefore independent from our MOSFIRE data and our own spectral fits.
The age--color relation is clearly detected in the first panel. The distribution of sSFR (averaged over the last 100 Myr) also shows a clear trend, but one that is perpendicular to that of the stellar ages. Finally, the dust extinction seems to be higher in both younger and more star forming galaxies. In Section~\ref{sec:compare_sfh} we demonstrated that these three quantities do not change significantly when adopting a different SFH model, therefore the results shown in Figure~\ref{fig:mosfire_UVJ} are robust. These results are qualitatively consistent with our study at lower redshift \citep{belli15}.


\section{Assessing the Impact of Merging on the Number Density Evolution of Quiescent Galaxies}
\label{sec:appendix_merging}

In Section~\ref{sec:numdens} we inferred the rate at which massive galaxies are quenched, by measuring the number density evolution of the population of galaxies that are quiescent and more massive than $10^{10.8}$\Msun, which we will call the \emph{reference population}. In our derivation we neglected, however, the contribution from galaxy merging, which has two distinct and opposite effects on the number density of the reference population: 1) when two galaxies belonging to the reference population merge together, the number density decreases; 2) when less-massive quiescent galaxies increase their stellar mass via gas-poor merging and cross the threshold of $10^{10.8}$\Msun, thus entering the reference population, the number density increases.

A realistic modeling of merging and its effect on the number density evolution is outside the scope of the present work. However, we can obtain an approximate estimate of the magnitude of this effect by making the following simplifying assumptions:
\begin{itemize}
  \item We only consider major mergers, defined as having a mass ratio larger than 1:4. This is justified given that minor mergers contribute four times less than major mergers to the mass growth of massive galaxies \citep{man16}.
  \item Mergers involve a range of mass ratios, but to simplify the calculations we only consider a fixed mass ratio value. Observations of pair fractions showed that the distribution of mass ratios at high redshift is virtually flat in logarithmic space \citep{newman12}, consistent with theoretical predictions \citep{hopkins10mergers}. Using this distribution it is then possible to calculate the average value, which for major mergers is 1:1.85, corresponding to a mass growth for the central galaxy of $\Delta \log \Mstar \approx 0.2$. We therefore assume that in each major merger, the \emph{central} galaxy increases its stellar mass by this amount, while the \emph{satellite} galaxy disappears.
  \item There are substantial uncertainties on the measurement of merger rates. Following the results of \citet{newman12}, we assume a major merger fraction $f_\mathrm{mg} = 10\%$, constant with redshift and stellar mass, and an effective merger timescale $\tau_\mathrm{mg} = 1$ Gyr. This means that at any given moment, 10\% of galaxies are observed to be in a major merger, and each merger is visible for 1 Gyr.
  \item The mass function of quiescent galaxies, $\Phi_\mathrm{qsc}$, in this mass and redshift range can be approximated fairly well by a Schechter function with fixed slope $\alpha = -0.4$ and characteristic mass $M^\ast=10^{10.8}\Msun$, and normalization that varies with redshift \citep{muzzin13}.
\end{itemize}

We can now calculate the contribution of merging to the number density evolution of quiescent galaxies. We start by defining the (redshift-dependent) number density of the reference population:
\begin{equation}
\tilde \Phi \equiv \int_{10.8}^{+\infty} \Phi_\mathrm{qsc}(\log M) \, d \log M \; .
\end{equation}
This number density decreases every time two of the objects in the reference population merge together. In our simple model, this happens every time the satellite galaxy is above the mass threshold of $10^{10.8}\Msun$, which means that the central must be above $10^{11}\Msun$. The number density of galaxies leaving the reference population over a timescale $\tau_\mathrm{mg}$ is then:
\begin{equation}
f_\mathrm{mg} \cdot \int_{11.0}^{+\infty} \Phi_\mathrm{qsc}(\log M) \, d \log M = 0.49 \, \tilde \Phi \, f_\mathrm{mg} \; .
\end{equation}
This is actually an upper limit, because it includes cases in which a massive quiescent galaxy merges with a star-forming satellite.

At the same time, the reference population will also grow due to less massive objects that cross the mass threshold. In order for this to happen, the central must have a mass in the range $10^{10.6}\Msun < \Mstar < 10^{10.8}\Msun$, and be quiescent, in order for this not to count as a quenching event. The number density of the reference population will then increase by:
\begin{equation}
f_\mathrm{mg} \cdot \int_{10.6}^{10.8} \Phi_\mathrm{qsc}(\log M) \, d \log M = 0.61 \, \tilde \Phi \, f_\mathrm{mg} \; .
\end{equation}
We can then calculate the \emph{rate} at which the reference population number density changes because of these two effects:
\begin{equation}
- \frac{0.49 \, \tilde \Phi \, f_\mathrm{mg}}{\tau_\mathrm{mg}} + \frac{0.61 \, \tilde \Phi \, f_\mathrm{mg}}{\tau_\mathrm{mg}} = 0.12\, \tilde \Phi \, \frac{f_\mathrm{mg}}{\tau_\mathrm{mg}} \; .
\end{equation}
If we use our estimate of $\tilde \Phi$ (shown in the top panel of Figure~\ref{fig:uvista_numden} and listed in Table~\ref{tab:numden}), we obtain $0.3 \cdot 10^{-6} \mathrm{Mpc}^{-3} \mathrm{Gyr}^{-1}$ at $z\sim2.5$ and $2 \cdot 10^{-6} \mathrm{Mpc}^{-3} \mathrm{Gyr}^{-1}$ at $z\sim1$.
These values are about two orders of magnitude below the observed growth of the quiescent population shown in the bottom panel of Figure~\ref{fig:uvista_numden}.
Even allowing for the substantial uncertainties in our assumptions (such as the merger timescale, merger rate, and mean mass ratio), it is clear that merging has a much smaller effect than quenching on the evolution of the number density of quiescent galaxies at $z>1$.

We point out that the two opposite effects of merging on the number density have similar magnitude, and therefore almost cancel out, because the mass threshold chosen for our study happens to match the knee of the mass function. However, the absolute value of each of these two competing effects is small, $(1.5 - 10) \cdot 10^{-6} \mathrm{Mpc}^{-3} \mathrm{Gyr}^{-1}$, and not sufficient to significantly affect our results. We thus conclude that merging can be safely neglected in our analysis.

\end{document}